\documentclass[review]{elsarticle}
\RequirePackage{amsmath}
\usepackage[table,xcdraw]{xcolor}
\usepackage{lineno,hyperref}
\modulolinenumbers[5]

\usepackage{geometry}
\journal{Journal of Computers \& Security}

\usepackage{color}

\usepackage{times}
\usepackage{soul}
\usepackage{url}
\usepackage[utf8]{inputenc}
\usepackage{bbding}
\usepackage{graphicx}
\usepackage{booktabs}
\usepackage{algorithm}
\usepackage{algorithmic}
\usepackage{pifont}
\usepackage{graphicx}
\usepackage{subfigure}
\usepackage{multirow}
\usepackage{amsfonts}
\usepackage{color}
\newcommand{\fm}[1]{\textcolor{black}{#1}}

\begin{document}

\begin{frontmatter}

\title{FedRight: An Effective Model Copyright Protection for Federated Learning}

\author[UU,UI]{Jinyin Chen}
\ead{chenjinyin@zjut.edu.cn}
\author[UI]{Mingjun Li}
\ead{limingjuns@outlook.com}
\author[UT]{Yao Cheng}
\ead{yao.cheng@tuvsud.com}
\author[UU,UI]{Haibin~Zheng\corref{cor1}}
\ead{haibinzheng320@gmail.com}

\cortext[cor1]{Corresponding author.}

\address[UU]{Institute of Cyberspace Security, Zhejiang University of Technology, Hangzhou, China}
\address[UI]{College of Information Engineering, Zhejiang University of Technology, Hangzhou, China}
\address[UT]{Digital Service, TÜV SÜD Asia Pacific Pte. Ltd., Singapore}

\begin{abstract}
Federated learning (FL), an effective distributed machine learning framework, implements model training and meanwhile protects local data privacy.
It has been applied to a broad variety of practical areas due to their great performance and appreciable profits. 
\emph{Who really owns the model, and how to protect the copyright} has become a real problem.
% Unfortunately, since the first moment that the FL is brought up, the server and the client do not shake hands on the global model's property rights, not even to mention a good way to protect the copyright of the well-trained model.
Intuitively, the existing property rights protection methods in centralized scenarios (e.g., watermark embedding and model fingerprints) are possible solutions for FL.
%However, the embedding of the watermark will sacrifice the original predictive ability of the model, and the model fingerprints are easy to be removed by adversarial retraining. 
But they are still challenged by the distributed nature of FL in aspects of the no data sharing, parameter aggregation, and federated training settings. 
For the first time we formalize the problem of copyright protection for FL, and propose FedRight to protect model copyright based on model fingerprints, i.e., extracting model features by generating adversarial examples as model fingerprints. FedRight outperforms previous works in four key aspects: 
(i) \emph{Validity} - it extracts model features to generate transferable fingerprints to train a detector to verify the copyright of the model.
(ii) \emph{Fidelity} - it is with imperceptible impact on the federated training, thus promises good main task performance.
(iii) \emph{Robustness} - it is empirically robust against malicious attack on copyright protection, i.e., fine-tuning, model pruning and adaptive attacks.
(iv) \emph{Black-box} - it is valid in black-box forensic scenario where only application programming interface calls to the model are available.
Extensive evaluations across 3 datasets and 9 model structures demonstrate FedRight’s superior fidelity, validity and robustness.
% FedRight effectively realized the property right declaration and property right verification.
% The code of FedRight is open-sourced at~\url{https://github.com/research-limingjun/FedRight}.

\end{abstract}

\begin{keyword}
Copyright protection \sep federated learning \sep model fingerprints \sep robustness \sep black-box fingerprints 
\end{keyword}

\end{frontmatter}

% \linenumbers

\section{Introduction}
Federated learning \cite{mcmahan2017communication,mcmahan2016federated,yang2019federated,Justicia21,fd,journals/isci/JiangXZ21,WangZLZL21} is an emerging distributed learning framework for user data privacy protection.
In the federated learning scenario, there are usually one server and multiple clients. 
Depending on the specific application, it can be divided into horizontal federated learning, vertical federated learning and transferable federated learning.
Horizontal federated learning (HFL) is the most popular one among them.
The client trains a local model with locally collected data, and then only uploads the model information to the server.
The server generates a global model with aggregation algorithms by aggregating the model information uploaded by clients, and distributes the global model back to the client for further training in an iterative manner.
The raw training data are safely kept in the local client during the whole training process. 
Thus the FL successfully maintains the privacy of clients' training data, meanwhile implements the distributed training whose accuracy is comparable to that of the centralized training. 
Due to its privacy-preserving nature and efficient distributed training mechanism, FL has been widely applied to practical scenarios such as bank loans~\cite{conf/icdm/Shingi20}, medical diagnosis~\cite{KuoP22}, and recommendation systems~\cite{rs}.

% Unfortunately, ever since FL is proposed, the server and the client do not shake hands on the global model’s copyrights~\cite{xue2020intellectual,Boenisch21}, not even to mention a good way to protect the copyright of the well-trained model. 
Unfortunately, there lacks a proper method to protect the copyright of the FL-trained model.
This problem directly leads to the difficulty for the owner of the model to claim the copyright even if the model is suspected to be illegally used by others for profit.
% It poses a challenge to the security of FL. 
How to better protect the property rights of the model and interests of the participants is an urgent issue that needs to be solved.

% Different from centralized training, under the distributed mechanism of FL, the cooperative training of multi-clients and trusted servers makes it necessary to discuss who is responsible for the Intellectual Property (IP) rights declaration before discussing specific protection methods.
% We consider that since the difficulty of defining whether clients are trustworthy while leaving the IP rights declaration to the client, it is inevitable to change the settings of federated training.
% Therefore, this paper examines model copyright protection techniques that revolve around the trusted servers to perform this operation.

There are a number of existing methods for model copyright protection in the centralized scenario.
% Regarding specific property rights protection techniques, they are widely used in the centralized scenarios. 
Watermark embedding ~\cite{UchidaNSS17,Vybornova21,LiZZDZX20,GuoP18,LiHZG19, waffle} and model fingerprinting~\cite{ZhaoHLMCH20,LukasZK21,CaoJG21} are the two mainstream model ownership verifications.
They require the model owner to embed a unique secret watermark or generate a unique secret fingerprint before releasing the model. 
When it comes to the situation where the model owner needs to claim the ownership of the model, the owner can use the secret watermark or fingerprint in combination with the behavior of the suspect model to prove so. 
However, embedding the watermark needs to change the model parameters, the process inevitably affects the performance and efficiency of the main task of federated training.
Model fingerprints methods extract model features (e.g., gradients) to generate an adversarial example~\cite{LuoLWX18,HuangPGDA17,ZhengZGLP20} and use the transferability of the generated adversarial examples to determine whether the use of the suspect model is unauthorized. However, fingerprints are easily erased by adversarial retraining~\cite{LukasZK21}.
% In a word, model copyright protection is still in its infancy, and it is challenging towards model copyright protection in FL.

Different from the centralized training, FL is distributed and involved different parties such as the server and the clients. 
However, according the assumption of the FL, a client may not be trustworthy but a server is, i.e., a trusted and honest server in FL.  
Therefore, it is more rational to rely the copyright protection on the server side than on the client side. 
However, there are still challenges to directly migrate the centralized copyright protection to FL.

We summarize the main challenges addressed for the model copyright protection in FL:
(i) \emph{Data limitation} - An honest server cannot access the training data owned by the clients to make watermarks or model fingerprints.
(ii) \emph{Accuracy sacrifice} - The protection should not excessively sacrifice the server model's accuracy, as well as the FL's training efficiency.
(iii) \emph{Malicious attacks} - The protection method at the server side needs to deal with downstream attacks, i.e. fine-tuning, model pruning, etc., as well as malicious clients upstream in collaborative training.
(iv) \emph{Black-box ownership verification} - The model copyright verification should be practical for black-box ownership verification scenario (i.e., without any knowledge of the model's architecture or parameters), for example, only Application Programming Interface (API) calls to query the suspect model are available.

To address these challenges, we propose FedRight, the first approach proposing to use model fingerprints to protect model copyright in FL.
FedRight is  based on model fingerprints techniques but with improved robust against adversarial retraining attacks.
Generally, we rely on the honest FL server to generate fingerprints for the FL-trained model and train a separate model using these fingerprints and their outputs. 
Specifically, by using secret key samples, the server leverages on the extracted global model features and generates a set of adversarial examples as model fingerprints.
Note that the key samples here are not necessarily of the same distribution with the training or testing samples. 
Then we use these fingerprints as input to the model and obtain the feature distributions of these fingerprints. 
The feature distribution of the key samples is used to train a new model, i.e., the detector, which is to predict the ground-truth class encode of the key samples.
When testing a suspect model, we obtain the feature distribution of the key samples output by the suspect model, and use the detector to predict the key samples' ground-truth class. 
When the accuracy of the detector exceeds a threshold, the ownership of the model is claimed.

Moreover, FedRight has designed the model fingerprint to have adaptive enhancement capability. 
It gradually adds model features in response to the changes in the global model during federation training. 
During the verification phase, FedRight only accesses the model's output of key samples, which is suitable for the black-box forensic scenarios.
We conducted extensive experiments to evaluate the validity, fidelity, efficiency and robustness of FedRight.

The main contributions of this paper are summarized as follows.
\begin{itemize}
\item 
We propose the FL-oriented model copyright protection method - FedRight by relying on the honest server to generate robust fingerprints without any knowledge of the training data on the client side. 

\item 
FedRight innovatively introduces a detector to capture the relationship between the feature distribution of the key samples output by the target model and the key samples' label. FedRight can effectively verify the ownership of the target model according to the accuracy of the detector, i.e., measuring to what extent a suspect model's behavior on these key samples aligns exactly with the target model.
 
\item 
We conducted extensive experiments to evaluate FedRight on 3 datasets and 9 models. The experimental results show the advantages of FedRight compared to previous work and satisfies validity, fidelity, robustness, and black-box ownership verification in FL scenarios.
\end{itemize}

\section{Related Works\label{RW}}

\subsection{Centralized Model IP Protection}
Existing means of intellectual property (IP) protection are mainly applied to the deep neural network (DNN) in centralized scenarios.
There are two streams of IP protection methods, i.e., watermarking and fingerprinting. 
Since Uchida et al.~\cite{UchidaNSS17} first proposed a watermarking model framework in a white-box scenario.
It gets further upgraded in the face of the restrictions on access rights in the black-box scenario.
Li et al.~\cite{WeiLLL018} combine common data samples with exclusive ``logos" and train models to predict them as specific labels so that the ownership of the model can be verified by a third party.
Jia et al.~\cite{JiaCCP21} propose entangled watermark embedding to address watermark removal attacks. 
In order to be robust against copyright evasion attacks, Zheng Li et al.~\cite{LiHZG19} propose a blind watermarking method to generate key samples with a distribution similar to the original samples.
% Watermarking techniques are effectively used in DNNs for IP protection. 
However, embedding watermarks changes the model parameters and this process inevitably affects the performance of the main task of the model.
% The use of model fingerprints to identify stolen models is a novel means of IP rights protection.
Fingerprinting is another IP protection that does not change the model parameters.
Zhao~\cite{ZhaoHLMCH20}, Lukas et al.~\cite{LukasZK21} used adversarial examples as model fingerprints and exploited their transferability to verify model IP rights.
Fingerprinting methods generate model fingerprints by extracting model features, hence there is no impact on the model performance.
However, model fingerprints can be removed in the face of adaptive adversarial retraining attacks. Moreover, the privacy-preserving nature of FL, where there is no training data available, poses great challenge to the fingerprinting process on the server.

\subsection{IP Protection in FL}
% IP protection of federal models, Tekgul~\cite{waffle}, Fang-Qi Li~\cite{ms}, and Bowen Li et al.~\cite{fedipr} designed corresponding watermarking techniques to solve this problem.
The mainstream IP protection in FL is still based on watermarking~\cite{waffle}\cite{fedipr}\cite{ms}. 
Specifically, Tekgul et al.~\cite{waffle} proposed WAFFLE which achieves IP protection by embedding watermarks on the server. 
Bowen Li et al.~\cite{fedipr} proposed FedIPR which embeds and detects watermarks by each client independently.
Fang et al. ~\cite{ms} designed the Merkle-Sign watermarking framework, which combines the state-of-the-art watermarking scheme and a security mechanism designed for distributed storage to protect both privacy and ownership. 
% However, how to better combine security mechanisms and watermarking schemes remains to be investigated.
However, they either surfer the inherent limitation of the watermarking scheme or pose changes to the training of federated learning, both of which lead to negative impact on the model performance. 

\subsection{Attacks against IP Protection}
Malicious attacks may launch attacks aiming at obtaining the model without degrading its accuracy, and meanwhile preventing the model owner from proving his/her ownership.
There are model modification methods~\cite{UchidaNSS17}\cite{RouhaniCK19}\cite{NambaS19}, copyright evasion attacks~\cite{HitajHM19}, and removal attacks~\cite{removal}.
The model modification includes model fine-tuning~\cite{UchidaNSS17}, model pruning~\cite{RouhaniCK19}, model compression~\cite{UchidaNSS17}, and model retraining~\cite{NambaS19}.
In order to evade the legitimate owner's verification, in the copyright evasion attack against IP protection~\cite{HitajHM19}, the attacker will try to construct a detector to detect whether the queried sample is a clean or possibly critical sample. Once the detector determines that the queried instance is a possible critical sample, the stolen model will return a random label.
Removal attacks, the attackers attempt to remove the watermark. Shafieinejad et al.~\cite{rm1} studied removal attacks based on backdoor watermarking schemes and proposed a method to detect whether a model contains a watermark.
Wang et al.~\cite{rm2} used generative adversarial network (GAN) to detect and reverse the backdoor trigger in the model and then fine-tune the model with the reversed trigger to remove the backdoor based watermark.
Chattopadhyay et al.~\cite{rm3} used GAN to generate samples for retraining that can obtain a model with similar performance while removing the watermark.

\section{Preliminaries and Background}
In this section, we introduce the horizontal federation framework and the background knowledge for generating model fingerprints.

\subsection{Horizontal Federated Learning}
Horizontal federated learning~\cite{hfl} is applied to scenarios where the datasets of each client have the same feature space and different sample spaces. All private data are on the client and cannot be accessed by other clients. After each client $C_i$ performs model training locally, it uploads the model parameter $w_i$ to the server. 
Then, the server performs an aggregation operation on the uploaded parameters to form a global model parameter $w_{g}$, which is then returned to each client to continue training.
The commonly used aggregation rule is as follows:

\begin{equation}\label{juhe}
w_{g}^{}= \frac{1}{K}\sum_{i=1}^Kw^{}_i
\end{equation}
where $K$ is the total number of clients participating in training.

% This process meets the needs of user privacy protection and data security.
In HFL, only information about locally trained model is shared. It thus ensures the privacy of the client's local data. 

\subsection{Key Samples}
% Extracting model features needs to be added to key samples to form model fingerprints. 
Key samples are the seeds to generate model fingerprints. 
Key samples are normally protected from attackers' access. 
Existing model protection methods generate fingerprints using key samples from training data. 
However, this could be an issue in FL since the server in FL does not have access to client data, so it cannot use the training data as key samples.
If the key samples are part of the training data, attackers are prone to use the adversarial retraining attack of training data to eliminate fingerprints. 
Therefore, the key samples used in this paper are independent of the training data. 
This could 1) fits in the FL scenario where training data are not available; and 2) greatly affects the main task performance if the attacker performs retraining attack.  
% In contrast, using special key samples is not easy to be stolen, and even if it is stolen, the retraining process can greatly affect the main performance since it is not related to the training data.

Using non-training data as key samples can also generate valid model fingerprints because FedRight's model fingerprints consist of key samples and model-specific features. Specifically, the key sample is only the carrier of the model fingerprint, which depends critically on the extracted model-specific features. Thus, FedRight proposes an effective method for generating fingerprints without training data.

\subsection{Adversarial Examples}
Existing work~\cite{ZhaoHLMCH20,LukasZK21} utilizes adversarial examples as model fingerprints.
Model can easily be fooled by well-designed adversarial example $x_{adv}$, i.e., by adding small perturbations $\sigma$ to the normal example $x$.
There are many algorithms for generating adversarial examples, such as FGSM~\cite{fgsm}, C$\&$W~\cite{cw}, PGD~\cite{pgd}, etc.
Given a sample $x$ and an output label $y_{orig}$, the attacker can always find well-targeted adversarial examples such that the output label is not $y_{orig}$, depending on the specific optimization process that can be divided into targeted and untargeted attacks.

\begin{equation}
 \begin{cases}
  & f(x+\sigma) \ne y_{orig}~~~, ~~~ untargeted~attack  \\
  & f(x+\sigma) = y_{target}~, ~~~ targeted~attack
\end{cases}
\end{equation}
where $f(\cdot )$ is the input to the target model; $y_{target}$ denotes the expected target label.

Since adversarial examples are generated based on the model, e.g., model structure and parameters, they can be effectively used as fingerprints to identify a model. 
As long as the attackers are not aware of the adversarial examples, especially the corresponding model output of these adversarial examples, model owners can verify the ownership of the model by testing the suspect model with the secret adversarial examples and the corresponding predictions.
% This feature is also used to fool attackers because they only know the labels of the training data. 
% Thus, the exploit of adversarial examples can be used to determine whether a suspect model is a pirated model.

% Since the adversarial perturbation is highly related to the structure and parameters of the model, it means that the use of adversarial examples can effectively be used as model fingerprints, exploiting their transferability to detect pirated models.

\subsection{Model Fingerprints}
Model fingerprinting is a method that produces a model fingerprint $F$ according to the target model to realize copyright protection.

It includes the following two algorithms:

\textbf{Generate model fingerprints.} $F = Generate(G;D_{key})$. The generation process has access to global model $G$ and key samples $D_{key}$, and uses this knowledge to generate the model fingerprint $F$ for a specific label $F_y$.
% verification keys $F_y=$$\left \{ G(x)|x\ \in F \right \}$.

\textbf{Validate model fingerprints.} $F^{pre}_y = Validate(G_{sus};F)$.
The model owner validates the suspect model $G_{sus}$ using the model fingerprint $F$, and the obtained output prediction label $F^{pre}_y$ are compared with $F_y$ to verify the ownership.

The evaluation of the suspect model ownership using the validation algorithm is based on an empirically determined threshold $\alpha$. 
% The threshold $\alpha$ is obtained from extensive validation tests performed on models with IP declaration and models with non-IP declaration.
If the matching rate between $F^{pre}_y$ and $F_y$ is greater than $\alpha$, it is verified as a stolen model. 
Since the model fingerprint, i.e., the adversarial example, has transferability, the model ownership can still be verified even though the suspect model is slightly modified from the original model.

Existing model fingerprint verification algorithms only using prediction labels are vulnerable to ambiguity attacks, where a fake owner uses samples with the same output label as the original model fingerprints to falsely claim the model ownership.
For example, the model owner is considered to have ownership of the model using a key sample output label ‘1’. However, the attacker uses a non-key sample output label also ‘1’, which is an eventuality and makes the ownership of the model confusing.
We take this into account when designing our fingerprint verification algorithm. 
We use the output distribution features of the model instead of prediction label only, which will be detailed in Section~\ref{sec:verification}.

\begin{figure}[t]
  \centering
  \includegraphics[width=0.7\textwidth]{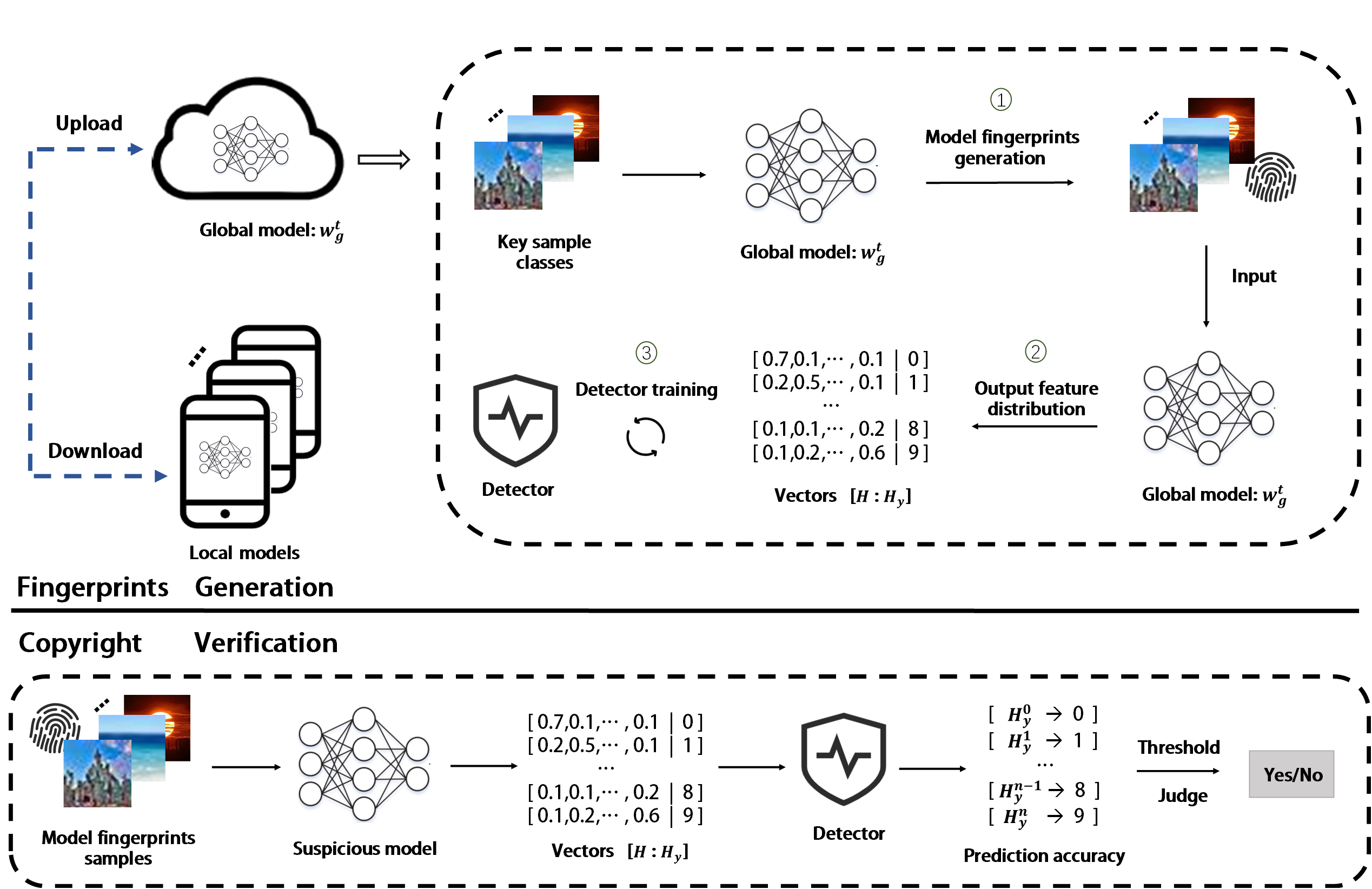}
  \caption{Overview of FedRight. The server clones a global model, extracts the inherent features of the model, and generates adversarial examples from key samples as model fingerprints. After input to the global model, the output distribution features are obtained and labeled to train the detector. Using the transferability of the adversarial example, model owners can use the detector to determine the ownership of the suspect model.  }
 \label{framework}
\end{figure}

\section{Threat Model}
This paper takes the server as the initiator of property protection, and the main threat target is the upstream and downstream attackers.

Suppose the horizontal federation contains $K$ clients $C_i$ and a trusted server, where each client has local data $D_i$, $i\in\left \{ 1, 2, ... K \right \}$. 
All clients collaborate to train a global model $G$, but the distributed mechanism brings uncertainties,
such as the existence of malicious client $C_j^m$ during the training process who intends to exploit the high-value global model $G$ for illegal profit.
Moreover, when the federation training is finished, the public sharing of the global model leads to a downstream attacker $P_m$ being able to tune it to avoid the owner's IP rights traceability, and then uses the tuned model for other illegal avenues. 
To clearly describe such a threat scenario, we give the formal definitions of attacker and defender as follows.

\textbf{Attacker ability.}
We suppose the attacker can be the upstream malicious client $C_j^m$ or the downstream malicious party $P_m$.
The upstream malicious client $C_j^m$ has the generally accessible information set
$<\Theta_g, L, l_r, R_{aggr}>$
in the FL system, where $\Theta _g$ is the global model parameter, $L$ is the loss function, $l_r$ is the learning rate and $R_{aggr}$ is the aggregation rule. 
However, $C_j^m$ cannot access or manipulate other clients' data $D_i$. The downstream malicious party $P_m$ can generally only have the global model parameter $\Theta_g$ information. 
But both attackers can use some modification attacks against the global model (e.g., model fine-tuning~\cite{UchidaNSS17}, model pruning~\cite{RouhaniCK19}), denoted as $\Theta_g'$=$Modi (\Theta_g)$, 
thereby making it different from the original model $G(\Theta_g)$ $\ne$ $G(\Theta_g')$, to prevent the trace verification of model IP protection, but still has high accuracy: 
$\left \|Acc(D_{test} ;G(\Theta_g'))-Acc(D_{test} ; G(\Theta_g) ) \right \|<\delta $, where $\delta $ is a small number.

\textbf{Defender ability.}
The goal of the defender $P_{d}$ is to protect the IP of models collaboratively trained by multiple clients in the FL scenario. 
In a practical scenario, the $P_{d}$ has no white-box access to the model stolen by the attacker. 
However, the $P_{d}$ should have black-box access to the suspect model $G_{sus}$,
i.e., it can query the suspicious model $G_{sus}$ with the $< X_{key} ,Y_{key}>$ and obtain the output $Y_{pre}$ $\gets$ 
$ f( X_{key},G_{sus})$ to verify the model's IP
$ Ver(< Y_{pre}= Y_{key}?>; G_{sus}) $.
A rational assumption would be 1) the defender has the white-box access to the model to be protected, since he/she owns the copyright; 2) the defender only has black-box access to the suspect model deployed by others.

% The goal of the defender is to property-protect the original model while being able to identify the stolen model deployed remotely by the attacker.
% During the property rights declaration process, the defender has white-box access to the global model. During the process of verifying model property rights, it has black-box access to the verified model.

\section{Methodology\label{Method}}

We rely on the server to generate fingerprints for the purpose the model copyright protection. 
An overview of FedRight is shown in Fig.\ref{framework}.
The overall process is consisted of two phases, i.e., fingerprints generation and copyright verification. Specifically, the first phase is divided into three steps:
\normalsize{\textcircled{\scriptsize{1}}}\normalsize~generate adversarial examples as model fingerprints;
\normalsize{\textcircled{\scriptsize{2}}}\normalsize~obtain the output distribution features of the model fingerprints and label them with specific labels;
\normalsize{\textcircled{\scriptsize{3}}}\normalsize~detector training. 
The verification phase happens when there is a suspect model and the model owner intends to verify the ownership of the model. 
The owner needs to obtain the feature distribution of the fingerprint samples by querying the suspect model. 
Then the obtained distribution is fed to our detector to see whether the prediction accuracy is high enough to claim the model's ownership.

\subsection{Model Fingerprints Generation}\label{zhiwen}

Since the server does not have access to the client's local data, but can explicitly know the specific training task. Based on this knowledge, we collect some key samples $D_{key}$ that are not related to the training data. 
For example, if the server knows that the main task of federated training is the Handwriting Digit Recognition, it can use face images as the key samples.
The class number of the key samples $D_{key}$ is also selected depending on the number of classes of the main task.
Usually, it is preferred to increase the number of key sample classes to improve the performance, e.g., with high detection rate and low false positive rate.

The selected key samples $D_{key}$ are used to generate model fingerprints $F$.
Specifically, 
by using adversarial attacks, key samples $D_{key}$ and extracted model perturbations are combined to form adversarial examples $D_{adv}$ as model fingerprints.
The goal of the adversarial attack is to minimally interfere with the normal examples while maximally misleading the classifier with a high confidence level. This can be modeled as an optimization problem. 
We introduce a generic model of adversarial attacks against a DNN, named $\rho$-loss, defined as:

\begin{equation}\label{adv}
\begin{split}
arg min\left \{ \epsilon \left \| \rho   \right \| +\lambda  Loss(y,f_{pre}(\Theta ,x)) \right \}~~~~~~~~~~~~~~~~~~~~~~~~~~~
\\
s.t.~\rho  = X_{adv} - X_{key} ~~,~~
X_{adv} \in \left \{ D_{adv} \right \}, X_{key} \in \left \{ D_{key} \right \} ~~,~~ y \in \left \{ y_{orig},y_{target} \right \}
\end{split}
\end{equation}

where $\epsilon$ is the scale factor used to balance the order-of-magnitude difference in the perturbation;
$\rho$ denotes the perturbation between the adversarial example $X_{adv}$ and the key example $X_{key}$. $X_{adv}$ is the sample in the set of adversarial examples $D_{adv}$, and $X_{key}$ is the sample in the set of key examples $D_{key}$;
$Loss(\cdot ,\cdot )$ is the loss function of the model; $f_{pre}(\cdot ,\cdot )$ is the prediction result of the model; $\Theta$ denotes the parameters of the model; $x$ and $y$ denote the input key example $D_{key}$ and the corresponding output class labels, respectively.
When $\lambda$=1 and $y=y_{target}$, $\rho$-loss represents a targeted attack, and when $\lambda$=-1 and $y=y_{orig}$, $\rho$-loss represents an untargeted attack.

The targeted attack has a clear optimization direction and better attack effect compared with the untargeted attack.
Thus, we use targeted attacks by setting $\lambda$=1,
and generate model fingerprints $F$ for the corresponding classes.
It is worth noting that the FedRight is generic framework transferable for other adversarial attacks including C\&W~\cite{cw}, PGD~\cite{pgd}, etc.

\begin{figure}[h]
\centering{
        \includegraphics[width=0.6\textwidth]{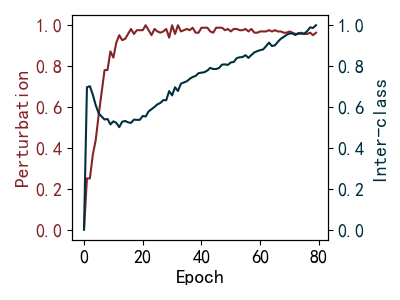} }

\caption{
The correlation between the model feature perturbations added and the distance between model fingerprint classes in the process of generating model fingerprints.
}\label{pd}
\end{figure}

To better explain the uniqueness of model fingerprints that can be exploited,
we visualize the inter-class distance of the fingerprints during the global model training as shown in Fig.\ref{pd}, and find that the training of the global model contains two phases, the oscillation phase $T$ (i.e., the first 20 epochs) and the linear growth phase (i.e., the last 60 epochs).
In the oscillation period $T$, the Mean Square Error (MSE) value of our proposed model fingerprint increases continuously with the training epoch, but the growth curve has a certain time delay $t_d$. During the linear growth, the MSE value of the model fingerprint is essentially constant.
During the oscillation period $T$, we infer that the model is rapidly learning the ``knowledge" in the data. Therefore, the model fingerprint is also rapidly collecting the features trained by the model, which is reflected by the increasing MSE value of the model fingerprint. The features of the model fingerprint are collected after the model learns the ``knowledge", so there is a certain time delay $t_d$.
In the linear growth process, the model just keeps converging the intra-class distance and increasing the inter-class distance. We infer that there are fewer features that can be provided to the model fingerprint at this time, and therefore the MSE value of the model fingerprint remains essentially unchanged.
To sum up, the model fingerprint we propose is directly related to the global model, for which the global model is unique.

\begin{figure}[h]
\centering
    \subfigure[The static model fingerprint generation]{
    \includegraphics[width=0.3\linewidth]{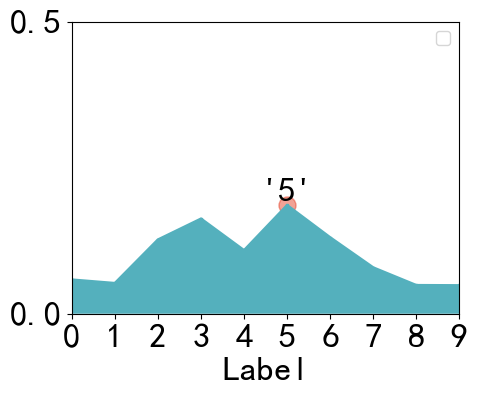}
    \includegraphics[width=0.3\linewidth]{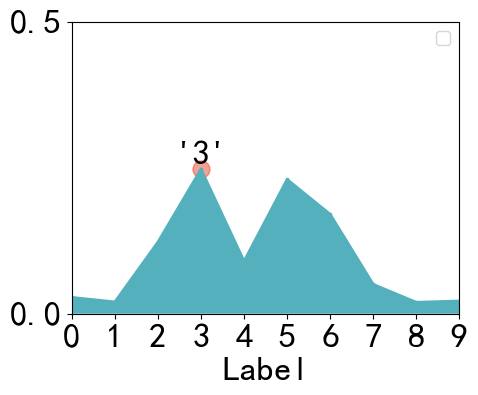}
    \includegraphics[width=0.3\linewidth]{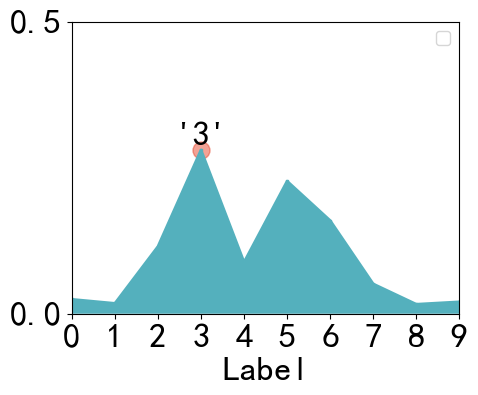}} \\
    \subfigure[\fm{ The adaptive enhancement to model fingerprint generation}]{
    \includegraphics[width=0.3\linewidth]{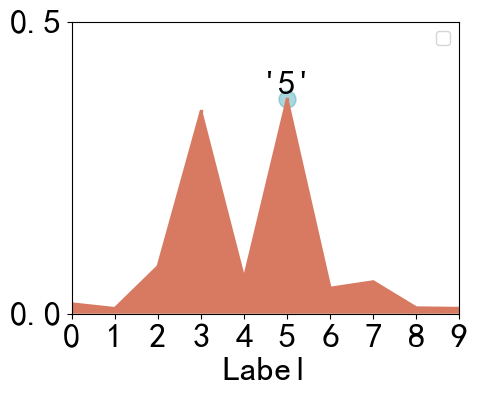}
    \includegraphics[width=0.3\linewidth]{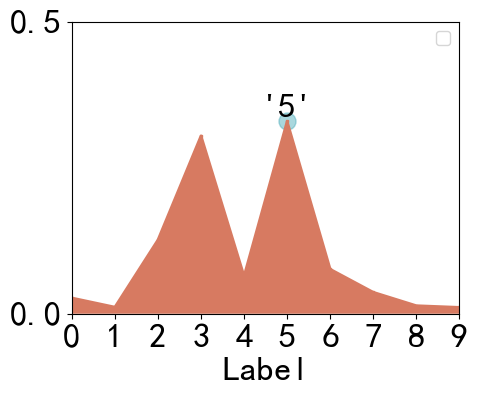}
    \includegraphics[width=0.3\linewidth]{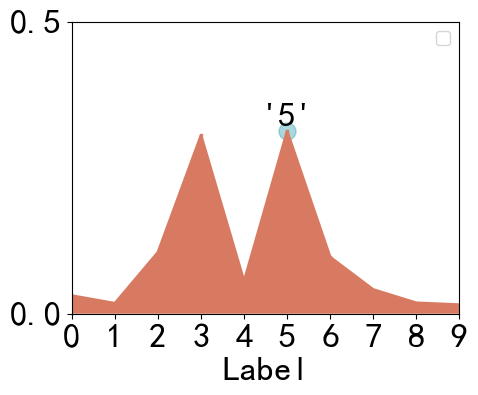}}
\caption{The output distribution feature vectors are visualized for a LeNet-5 model trained on MNIST dataset. The fingerprints are generated in the $50^{th}$ epoch and verified in the $50^{th}$, $51^{st}$ and $52^{nd}$ epochs. }\label{fb}
\end{figure}

Regarding the timing of model fingerprints generation, some options are possible, such as generating fingerprints during the training process or after the training.
If the generation of model fingerprints starts at the end of federated training, since the client gets the global model published by the server in each epoch of training, then it is possible for the malicious client to save the global models of previous epochs locally to evade IP verification. 
Therefore, the generation of model fingerprints needs to be performed during the federated training.
The other way is considering the model fingerprint generated from the beginning to the end of the whole federated training. It still brings some problems, not only to consider the time overhead of IP protection, but also at the early stage of training, the parameters of the global model change sharply with the iterative update in FL, thus the model fingerprint generated in the previous epoch to be input to the global model in the next epoch of training will lead the target label shifted, resulting in great change in the output distribution vector, i.e., the model fingerprint is invalid. 
As shown in Fig.\ref{fb}(a),
we randomly select the model fingerprint with target label `5' generated in epoch 50 and show the visualization of its output distribution vector in epochs 50 to 52.
It can be found that the model fingerprint generated in epoch 50 has been invalidated in the subsequent epochs.
It indicates that the static generation method is difficult to maintain the validity of fingerprints in dynamic training.

To solve this problem and ensure the stability of the model fingerprint, we propose a model fingerprint adaptive enhancement mechanism.

\subsection{Model Fingerprinting Adaptive Enhancement}

The model parameters usually change sharply at the early stage of FL training, and meanwhile, the initial model accuracy is low and not of high value for use. Therefore, fingerprint generation should be executed from the middle stage of training.

the middle stage of performing fingerprint generation is not absolute, but rather a relative range interval.
We visualized the global model parameter update rate and the timing of selecting fingerprint generation in 10 round intervals to get the set threshold for MNIST on LeNet-5 model as shown in Fig.~\ref{updaterate}.
For the model parameter update rate, it can be found that the global model parameter change rate increases rapidly in the first 5 epochs, decays rapidly from epochs 5 to 20, and gradually levels off afterwards. For the threshold $\alpha$, the $\alpha$ starts to gradually decrease in epochs 0-30 as the timing of selecting the execution of generating fingerprints shifts backward; after 30 epochs, the $\alpha$ starts to increase.
This indicates that in the preliminary model training, the initialized model parameters take some time to learn the features of the data, and the parameters are highly variable and not stable enough in this process, resulting in high $\alpha$. As the timing of generating fingerprints moved backward, the $\alpha$ began to gradually decrease. It is not until after 30 epochs that the $\alpha$ starts to increase, which may be due to the fact that as the overall number of epochs to generate fingerprints decreases, the amount of data generated is decreasing, which can affect the training of the detector $M$. It is worth noting that the overall $\alpha$ is not higher than 28\%, which is within the acceptable range.
To sum up, the fingerprint generation is performed starting in the middle of training having a lower threshold $\alpha$.

On the other hand, 
adaptive enhancement of fingerprints is achieved by setting a fingerprint validity detection checkpoint, which is defined as follows:

\begin{figure}[h]
\centering{
        \includegraphics[width=0.6\textwidth]{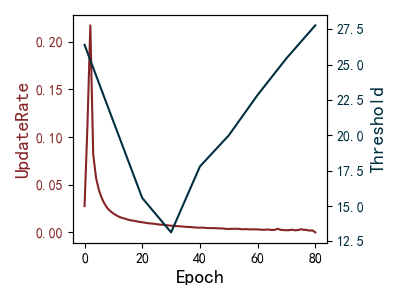} }

\caption{
The correlation between the update rate of the global model parameters and the threshold set at different time of performing the generation of fingerprints.
}\label{updaterate}
\end{figure}

\begin{equation}\label{en}
F^{t+1}= \begin{cases}
  & D_{adv}^{t} + \epsilon \left \| \rho   \right \|,~~~ Checkpoint(D_{adv}^{t})\times  \\
  & D_{adv}^{t}~~~~~~~~~~~~~~,~~~~  Checkpoint(D_{adv}^{t})\surd 
\end{cases}
\end{equation}
where $Checkpoint(\cdot)$ is to determine whether the target label of the adversarial sample has been shifted, i.e., whether the fingerprint is invalidated;
$\epsilon \left \| \rho   \right \|$ denotes adding the feature perturbation of the current epoch model;
The ``$\times $" indicates invalid,
and the ``$\surd$" indicates valid.

In each epoch, we use Algorithm 1 to judge whether the previous model fingerprint is valid and if it is invalid, we continue to add the model feature perturbation in the current epoch to achieve the effect of dynamic enhancement.

From Fig.\ref{fb}(b), we can also see that the model fingerprint generation algorithm with adaptive enhancement can maintain better across epochs, which effectively improves the validity of the generated fingerprints.

\subsection{Detector Training}\label{model}
The detector $M$ uses the MLP structure, whose training data
is constructed from the vector $H$ of feature distributions obtained from the model fingerprint $F$ input to the global model $G$. 
And $H$ is labeled with a specific label $H_y$ to train the detector $M$, where $H_y$ is consistent with the model fingerprint label, i.e., $H_y$ corresponds to the label of $D_{adv}$.

% where $H_y$ are consistent with the corresponding key samples $D_{key}$ classes.

\begin{equation}\label{H}
H = G( X_{adv} )~~ ,~~     X_{adv} \in \left \{ D_{adv} \right \} 
\end{equation}
% where $f(\cdot )$ denotes the input to the global model.

Training the detector $M$ using the output distribution feature $H$ effectively eliminate the risk arising from malicious clients using ambiguity attacks during the property rights verification process, i.e., using non-key samples consistent with the output labels of the key samples for property rights obfuscation. 
The reason is that the behavior of the distribution features $H$ of key samples and non-key samples are non-correlated.

Meanwhile, in order to improve the validation capability of the detector $M$, it can be trained by increasing the number of key sample classes $c$ to generate different classes of adversarial examples.

\begin{equation}\label{train}
Train(<H,H_y> ;M(w_m,c) )   
\end{equation}
where $Train(.;.)$ represents the training of the detector $M$ using the feature distribution vector $H$ and the given label $H_y$.
$w_m$ denotes the detector model parameters and $c$ denotes the number of classes of key samples.

The training of detector $M$ is independent of the main task FL training. Once the server aggregates the global model, it is immediately distributed to individual clients, while the current global model is cloned for generating model fingerprints $F$ and training the detector $M$.
This independence means that the copyright protection is not at the expense of model accuracy.

\subsection{IP Verification}
\label{sec:verification}

The trained detector $M$ can be used to identify the ownership of a suspect model.
In the verification phase, the model fingerprint is input to the suspect model and the output distribution vector $H$ is input to the detector $M$ to check whether the output label is consistent with the set label $H_y$.
Ownership of the model is determined based on whether the accuracy of the detector is above a set threshold $\alpha$.
In this process, only suspicious model inputs and outputs are utilized with no access to the internal parameters of the model.
The detail of FedRight is shown in Algorithm 1. 

\begin{center}
%\resizebox{0.8\textwidth}{!}{
\centering
\begin{tabular}{p{0.01\linewidth}p{0.9\linewidth}}
  \hline \hline
   & \textbf{Algorithm 1}: FedRight. \\
  \hline
   & \textbf{Input}:
   total number of clients $K$;
   dataset $\{D_i\}$ of each client,
   $i\in\{1,2,...,K\}$;  key samples $\{D_{key}\}$ in the server;
   the global epochs $t$;
   initial global model parameters $w_g^{t=0}$,  detector parameters $w_m^{t=0}$;
   hyperparameter $\epsilon=0.05$.\\
   & \textbf{Output}: the global model $G$ and detector $M$.\\
  \hline
  1. & Initialization: local model $w_{i}^{t=0} = w_g^{t=0}$.  \\
  2  &  \textbf{Role:}  Client $C_i$ \\
  3. &~~~~$w_i^{t+1} \Leftarrow Update(w_g^t)$\\
  4. &~~~~ {Local updates} ~$w_i^{t+1}$~{upload to Server}\\
  5. & \textbf{Role:} Server \\
  6. &~~~~~Calculate $w_g^{t+1}$ according to Eq.~(\ref{juhe}) \\
 7. & ~~~~~\textbf{Fingerprints generation:}\\
8.& ~~~~~~~~~~~Generating~$F$~according~to~Eq.~(\ref{adv}) \\ 
9.&~~~~~~~~~~ Get~the~output~vector~$H$~of~$F$~according~to~ Eq.~(\ref{H}) \\
10.&~~~~~~~~~~ Enhancing~$F$~according~to~Eq.(\ref{en})\\
11.&~~~~~~~~~~ Train~the~$M$~according~to~Eq.(\ref{train}) \\
 12. & \textbf{Return}: the global model~$G$ and  detector~$M$ \\

\hline \hline
\end{tabular}
%}
\end{center}

\subsection{Algorithm Complexity}\label{Complexity}
We analyse the complexity of FedRight on client-side and server-side, repectively.
On the client side, the time complexity is mainly dependent on the number of training epochs for the local model.
\begin{equation}
T_{client}  \sim  \mathcal{O}(t_l)
\end{equation}
where $t_l$ is the local training epochs.

On the server side, the server implements two main parts of work, and the two parts of work are independent of each other.
The first part of the work focuses on parameter aggregation of model parameters uploaded by clients and then distributed to each client.

\begin{equation}
Part_1: T_{server}  \sim  \mathcal{O}(K)
\end{equation}
where $K$ is the number of clients participating in the training.

The second part of the work focuses on the IP declaration process, i.e., the time complexity of the FedRight framework.
It consists of three parts: 
(1) generating the model fingerprints $F$;
(2) detecting whether the model fingerprints are invalid and performs enhancements;
(3) training the detector $M$.

\begin{equation}
Part_2: T_{server}  \sim  \mathcal{O}(Iter \ast n) + \mathcal{O}(n+Iter\ast n_{inv}) + \mathcal{O}(t_m)
\end{equation}
where $Iter$ is the number of optimization iterations, $n$ is the number of key samples, $n_{inv}$ is the number of invalid model fingerprints  and $t_m$ is the number of training epochs of the detector $M$.

Basically, the overhead introduced by FedRight is only $Part_2: T_{server}$, which can be done in parallel with the FL main task.

\section{Experiments Design and Setup\label{Exp}}

\textbf{Platform}:
i7-7700K 4.20GHzx8 (CPU), TITAN Xp 12GiB x2 (GPU), 16GBx4 memory (DDR4), Ubuntu 16.04 (OS), Python 3.6, pytorch1.8.2.

\textbf{Datasets}: We evaluate FedRight on three datasets,
i.e., MNIST~\cite{mnist}, CIFAR-10~\cite{cifar10} and CIFAR-100~\cite{cifar100}.
MNIST dataset contains 70,000 real-world handwritten images with digits ranging from 0 to 9.
Both the CIFAR-10 and CIFAR-100 datasets contain 60,000 color images of size 32 x 32, with 10 classes of 6,000 images each for CIFAR-10 and 100 classes of 600 images each for CIFAR-100.
The detailed information of datasets is shown in Table~\ref{table1}.

\begin{table}[h]
\centering
\caption{Dataset and model parameter settings.}\label{table1}
\resizebox{\linewidth}{!}{\huge
\begin{tabular}{ccccccccc}
\hline\hline
% \toprule
\textbf{Datasets}       &\textbf{Samples} &\textbf{Dimensions}  &\textbf{Classes}   & \textbf{Models}    &\textbf{Learning Rate}      &\textbf{Momentum}    &\textbf{Epoches} &\textbf{Bach Size}         \\ \hline
MNIST      & 70,000  &28$\times$28   & 10        & LeNet-1, LeNet-4, LeNet-5   & 0.005    &0.0001   &80      & 32           \\
CIFAR-10   & 60,000  &32$\times$32   & 10       & VGG-11, VGG-13, VGG-16  & 0.01  &0.9     &100      & 32         \\
CIFAR-100      &60,000  &32$\times$32    &100          &ResNet-18, ResNet-34, ResNet-50   &0.001   &0.001        &100      & 32         \\
\hline\hline
\end{tabular}}
\end{table}

\textbf{Number of clients}: 
We adopt 5 clients for FL training in all experiments except the parameter sensitivity experiments with different client numbers.

\textbf{Models}:
A number of classifiers are used for verification on various datasets.
For MNIST, LeNet-1, LeNet-4 and LeNet-5~\cite{LeNet5} are used for classification.
For more more complex image datasets,
CIFAR-10 and CIFAR-100,
VGG-11, VGG-13, VGG-16~\cite{vgg16}, and ResNet-18~\cite{resnet18}, ResNet-34, ResNet-50 are adopted, respectively.
Please refer to Table \ref{table1} for the above specific parameter settings.
All evaluation results are the average of ten runs under the same setting.

\textbf{Adversarial Attacks}:
In the main experiments a targeted attack method based on FGSM is used. 
Meanwhile, in sensitivity analysis experiments, C\&W and PGD adversarial attack methods are adopted to illustrate the trasferability of FedRight.

\textbf{Hyper-Parameters}:
For all experiments, we set the hyperparameter $\epsilon=0.08$, and select the key sample of $c=10$.

\textbf{Attack Methods}: 
To assess the robustness of FedRight, we used two well-known model modification attacks: fine-tuning and model pruning, as well as one adaptive attack: adversarial retraining.
Attacks by malicious clients during training are also considered, and copyright evasion attacks are analyzed.

\emph{FtuningAtt} (Model Fine-tuning Attack~\cite{UchidaNSS17}): It is a classic approach, allowing an attacker to re-train the  model with comparable performance to the original model, but of different parameters.
In this paper, the model is fine-tuned by using 10\% of the data in the test set.

\emph{PruningAtt} (Model Pruning Attack~\cite{RouhaniCK19}): It is designed to cut down targeted parameters and to obtain a new model that is different from the original model but still has similar accuracy.
We use the pruning algorithm in \cite{RouhaniCK19}, setting a certain percentage of the parameters with the smallest absolute value to 0. The percentage is set between 30\% to 90\% with an interval of 10\%.

\emph{AdaptiveAtt} (Adaptive Attack): We consider the AdaptiveAtt to eliminate the fingerprints and perform adversarial retraining of the model according to the different knowledge possessed by the attacker.

\emph{CollabAtt} (Collaborative Attack): Malicious clients can collaborate, including multiple upstream clients collaborating and upstream and downstream malicious parties to combine to damage IP.

\emph{MaliClientAtt} (Malicious Client Attack): In comparison with the WAFFLE method, the malicious client is set up to perform removal attack~\cite{rm1} during the fingerprint generation phase to counteract IP protection.

\emph{CopEvaAtt} (Copyright Evasion Attack~\cite{HitajHM19}): To evade the legitimate owner’s verification, the attacker attempts to construct a detector to detect
whether the queried sample is a clean sample or possibly a fingerprinting sample. The attacker deliberately returns a random label if the detector detects any fingerprinting sample.

\textbf{Baselines}:
For the watermarking IP protection scheme under FL discussed in related work, we use WAFFLE~\cite{waffle}, which is the latest and also on the server side, as a baseline.
And we also compare FedRight$_{\rm D}$, which is with the adaptive enhancement to model fingerprinting, with the static generation model fingerprints method FedRight$_{\rm S}$ where the fingerprints are statically generated in each epoch.

\textbf{Evaluation Metrics}:
We analyzed FedRight's performance by measuring the following metrics. 
(1) Fidelity: side effects on the main classification tasks, i.e., the global model accuracy (GMC).
(2) Validity: whether ownership of the model can be successfully verified,i.e., detector accuracy (DMC).
(3) Robustness: resistance to attacks, i.e., detector accuracy after being attacked (DMC$_{\rm Att}$). The evaluation results regarding GMC, DMC and DMC$_{\rm Att}$ are all shown in percentage in this section.

\textbf{Threshold $\alpha$}:
We set the threshold $\alpha$ mainly based on the statistics of two experiments. 1) Counting the DMC of the property rights model and the DMC$_{\rm Att}$ in the face of various attacks. 2) Using different training strategies (including federated learning and single-machine training), the same dataset and model training to get a non-IP model, i.e., without making an IP statement on it, and statistics of its DMC and DMC$_{\rm Att}$.

\section{Evaluation and Analysis}
In this section, we assess FedRight's performance by answering the following five research questions (RQs).

\begin{itemize}

% \item \textbf{RQ1}: Does the effectiveness of FedRight's property rights protection validate the ownership of the model?

% \item \textbf{RQ2}: What is the fidelity of FedRight property protection and does it have a significant impact on main task performance?

% \item \textbf{RQ3}: Is FedRight more robust in the face of malicious parties' property rights attack methods?

% \item \textbf{RQ4}: What are the advantages of model fingerprints technology over watermark technology?

% \item \textbf{RQ5}: Is the effectiveness of property rights protection affected by parameter sensitivity?

\item \textbf{RQ1}: Can FedRight effectively be used to claim the ownership of a given DNN model?

\item \textbf{RQ2}: Regarding the fidelity of FedRight, what is its impact on main task performance?

\item \textbf{RQ3}: How robust is FedRight in the face of attacks against IP protection?

\item \textbf{RQ4}: What are the advantages of model fingerprinting over watermarking in FL setting?

\item \textbf{RQ5}: How is the parameter sensitivity of FedRight?

\end{itemize}

\subsection{RQ1: Validity}
In this section, we evaluate the effectiveness of FedRight.
The purpose is to measure whether we can successfully verify the ownership of the target model under the protection of FedRight. 
We test DMC between the IP models and the non-IP models(i.e., models that are not under our ownership). In this case, the non-IP models were centrally trained with the same training data and model structure as the IP models. The detector $M$ is used to validate these models and to evaluate the final DMC.

\begin{figure}[t]
\centering{
        \includegraphics[width=0.6\linewidth]{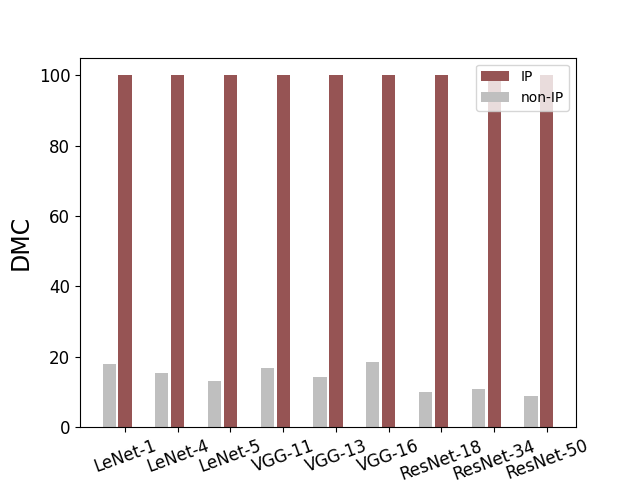} }

\caption{
The accuracy of detector for nine different models with three datasets. High DMC indicates the ownership of the model is verified.
}\label{validity}
\end{figure}

The results show that the detector effectively identifies the feature distribution of model fingerprints and predicts them to predefined labels with high accuracy.
As shown in Fig.\ref{validity}, all non-IP models only achieve an accuracy of 8.80\%-18.40\%, which is at about the same level of random guessing, i.e., the FedRight does not falsely claim ownership of non-IP models. 
In contrast, FedRight is encouragingly shown to achieve 100\% accuracy on models that declare IP rights. 
The reason is that each declared IP model has the unique model fingerprint, and using its output vector to train the detector can effectively predict it in the subsequent validation process with high accuracy to avoid misclassification.

\begin{center}
\fcolorbox{black}{white!20}{\parbox{0.97\linewidth}
{
\emph{\textbf{Answer to RQ1:}}
FedRight effectively verifies the ownership of the target model and achieves 100\% DMC without falsely claiming the ownership of the non-IP model.
}
}

\end{center}

\subsection{RQ2: Fidelity}
In this section, we evaluate the potential side effects of FedRight on the main task performance. 
\begin{figure}[h]
\centering{
        \includegraphics[width=0.6\linewidth]{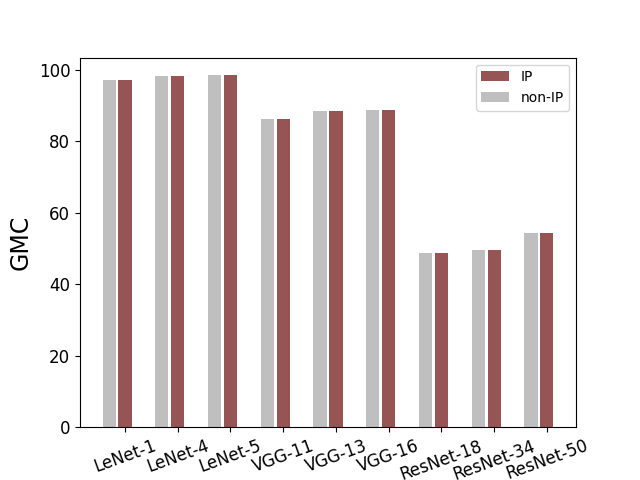} }

\caption{
The accuracy of the global model with IP protection and the global model for normal federated training.
}\label{fidelity}
\end{figure}
Fidelity requires that the IP protection are implemented without significant impact on the main task. 
We tested on three datasets with nine models to compare the difference in global model accuracy between normal federated training and federated training using the FedRight framework.

The experiments demonstrate that Fedright has excellent fidelity.
The results are shown in Fig.\ref{fidelity}, the global model that is with IP protection is consistent with the master task accuracy of the normally trained global model.
The reason is that FedRight and the main task training process are independent of each other. FedRight does not actively change any parameters of the global model.
It only uses the input and output of the model to realize the purpose of IP protection.

\begin{figure}[t]
\centering
    {
        ~\hspace{-0.08\linewidth}
        \subfigure[MNIST]{
            \includegraphics[width=0.35\linewidth]{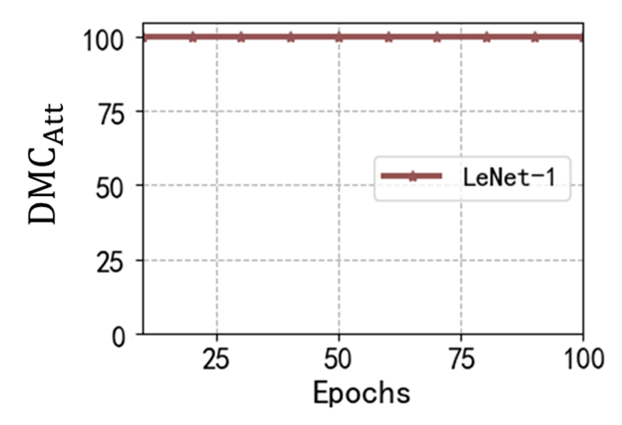} 
            \includegraphics[width=0.35\linewidth]{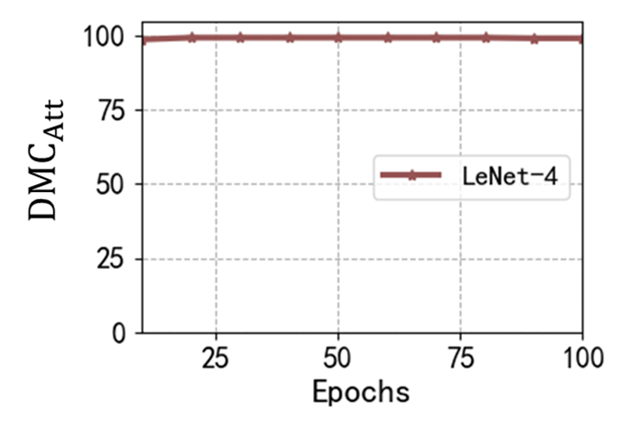} 
            \includegraphics[width=0.35\linewidth]{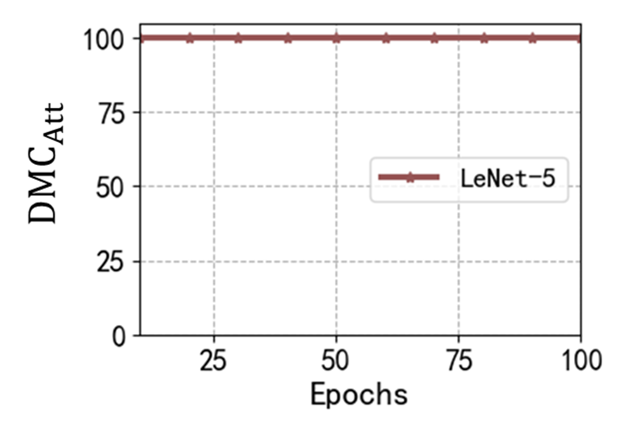}}
    } \\
    {
         ~\hspace{-0.08\linewidth}
         \subfigure[CIFAR-10]{
            \includegraphics[width=0.35\linewidth]{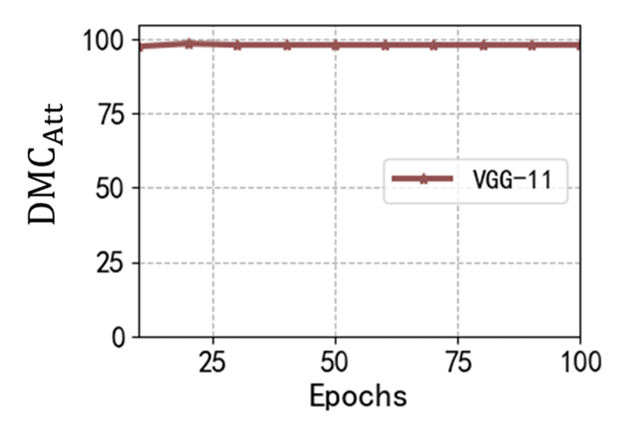}
            \includegraphics[width=0.35\linewidth]{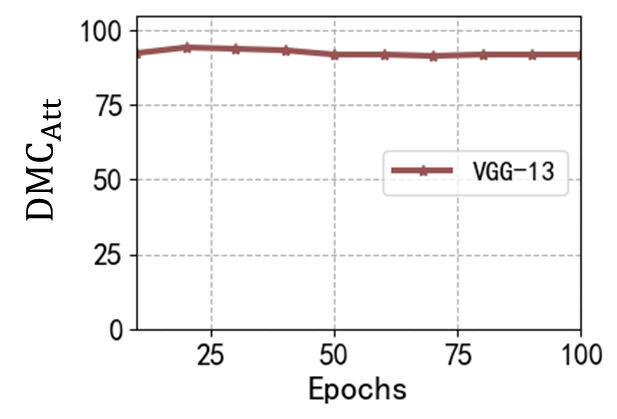}
            \includegraphics[width=0.35\linewidth]{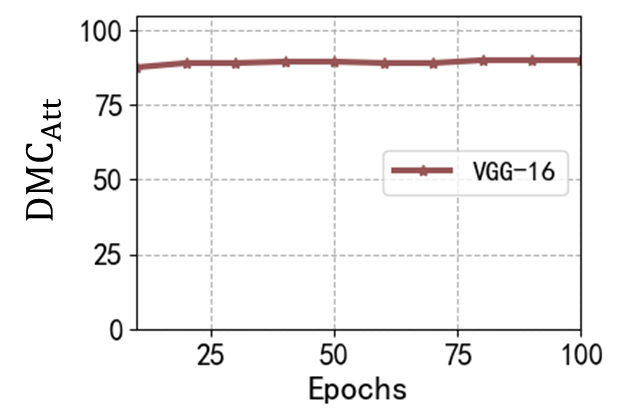}}
    } \\
    {
         ~\hspace{-0.08\linewidth}
         \subfigure[CIFAR-100]{
            \includegraphics[width=0.35\linewidth]{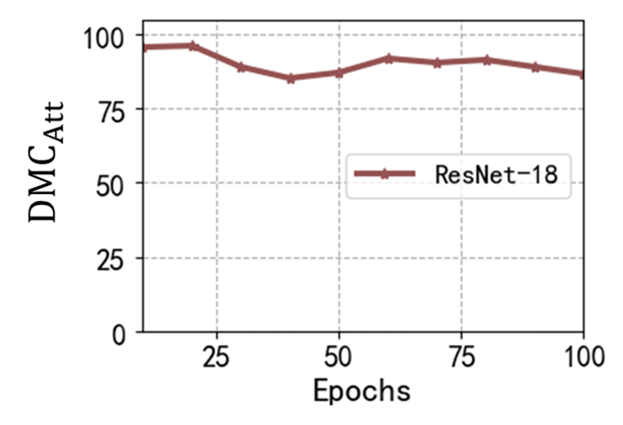}
            \includegraphics[width=0.35\linewidth]{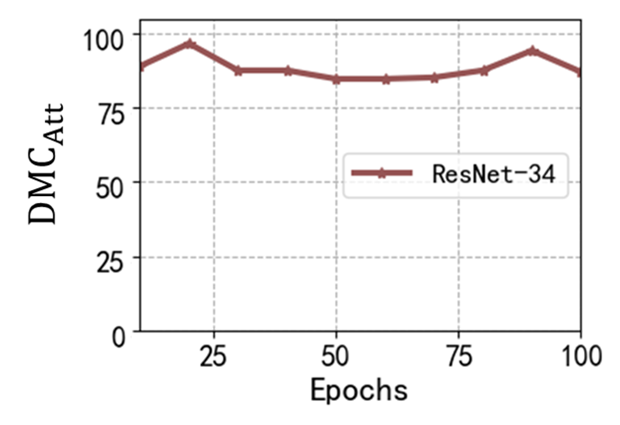}
            \includegraphics[width=0.35\linewidth]{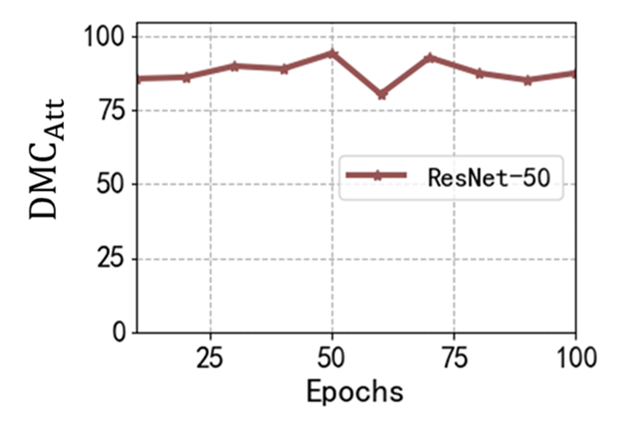}}
    }
   \caption{
The change in copyright verification accuracy in terms of DMC after performing fine-tuning using a 10\% test set for the nine models.
}\label{weitiao}
\end{figure}

\begin{center}
\fcolorbox{black}{white!20}{\parbox{0.97\linewidth}
{
\emph{\textbf{Answer to RQ2:}}
FedRight is independent of the main task training, thus it has a strong fidelity and does not have any side effects on the main task accuracy. 
}
}
\end{center}

\subsection{RQ3: Robustness}
In this section, we use FtuningAtt, PruningAtt and AdaptiveAtt to evaluate the robustness of FedRight, and also illustrate the CopEvaAtt.

\textbf{FtuningAtt.} FtuningAtt is a common strategy in practice, and we use 10\% of the data from the test set to fine-tune the trained model to measure the robustness of FedRight.
Fig.\ref{weitiao} shows that even after 100 epochs, FedRight still has the high accuracy of all models (only 19.50\% drop in the worst case). 
The reason behind this may be that FtuningAtt does not produce significant changes in the weights of the model.
Such modification that does not have significant side effects on the main task does not lead to a completely different model.

\begin{table}[h]
\centering
\caption{Changes in global model accuracy and detector model accuracy after performing PruningAtt on the 9 models.}\label{jianzhi}
\resizebox{0.9\textwidth}{!}{
\begin{tabular}{cccccccccccccl}
\hline\hline
\multicolumn{14}{c}{\textbf{MNIST}}                                                                                                                                                                                                                                        \\
\multicolumn{2}{c}{\multirow{2}{*}{\textbf{~~~~~~PruningAtt~~~~~~~~~}}} & \multicolumn{4}{c}{\textbf{LeNet-1 (Threshold=18.00\%)~~}}            & \multicolumn{4}{c}{\textbf{LeNet-4 (Threshold=15.50\%)~~}}            & \multicolumn{4}{c}{\textbf{LeNet-5 (Threshold=13.14\%)~~}}            \\
\multicolumn{2}{c}{}                                        & \multicolumn{2}{c}{\textbf{GMC}} & \multicolumn{2}{c}{\textbf{DMC$_{\rm Att}$}} & \multicolumn{2}{c}{\textbf{GMC}} & \multicolumn{2}{c}{\textbf{DMC$_{\rm Att}$}} & \multicolumn{2}{c}{\textbf{GMC}} & \multicolumn{2}{c}{\textbf{DMC$_{\rm Att}$}} \\ \hline
\multicolumn{2}{c}{\textbf{90\%}}                           & \multicolumn{2}{c}{50.33}        & \multicolumn{2}{c}{45.00}        & \multicolumn{2}{c}{93.53}        & \multicolumn{2}{c}{58.19}        & \multicolumn{2}{c}{72.39}        & \multicolumn{2}{c}{46.60}        \\
\multicolumn{2}{c}{\textbf{80\%}}                           & \multicolumn{2}{c}{80.53}        & \multicolumn{2}{c}{57.19}        & \multicolumn{2}{c}{97.33}        & \multicolumn{2}{c}{95.60}        & \multicolumn{2}{c}{90.79}        & \multicolumn{2}{c}{69.60}        \\
\multicolumn{2}{c}{\textbf{70\%}}                           & \multicolumn{2}{c}{91.18}        & \multicolumn{2}{c}{65.20}        & \multicolumn{2}{c}{97.32}        & \multicolumn{2}{c}{98.40}        & \multicolumn{2}{c}{92.36}        & \multicolumn{2}{c}{87.40}        \\
\multicolumn{2}{c}{\textbf{60\%}}                           & \multicolumn{2}{c}{94.35}        & \multicolumn{2}{c}{90.00}        & \multicolumn{2}{c}{97.63}        & \multicolumn{2}{c}{99.20}        & \multicolumn{2}{c}{96.65}        & \multicolumn{2}{c}{100.00}       \\
\multicolumn{2}{c}{\textbf{50\%}}                           & \multicolumn{2}{c}{95.37}        & \multicolumn{2}{c}{98.20}        & \multicolumn{2}{c}{97.98}        & \multicolumn{2}{c}{100.00}       & \multicolumn{2}{c}{96.84}        & \multicolumn{2}{c}{100.00}       \\
\multicolumn{2}{c}{\textbf{40\%}}                           & \multicolumn{2}{c}{95.87}        & \multicolumn{2}{c}{99.80}        & \multicolumn{2}{c}{98.18}        & \multicolumn{2}{c}{100.00}       & \multicolumn{2}{c}{98.01}        & \multicolumn{2}{c}{100.00}       \\
\multicolumn{2}{c}{\textbf{30\%}}                           & \multicolumn{2}{c}{96.15}        & \multicolumn{2}{c}{100.00}       & \multicolumn{2}{c}{98.28}        & \multicolumn{2}{c}{100.00}       & \multicolumn{2}{c}{98.46}        & \multicolumn{2}{c}{100.00}       \\ \hline\hline
                             % &                              &                 &                &                 &                &                 &                &                 &                &                 &                &                 &                \\ \hline\hline
\multicolumn{14}{c}{\textbf{CIFAR-10}}                                                                                                                                                                                                                                        \\
\multicolumn{2}{c}{\multirow{2}{*}{\textbf{PruningAtt}}} & \multicolumn{4}{c}{\textbf{VGG-11 (Threshold=16.90\%)}}             & \multicolumn{4}{c}{\textbf{VGG-13 (Threshold=14.20\%)}}             & \multicolumn{4}{c}{\textbf{VGG-16 (Threshold=18.40\%)}}             \\
\multicolumn{2}{c}{}                                        & \multicolumn{2}{c}{\textbf{GMC}} & \multicolumn{2}{c}{\textbf{DMC$_{\rm Att}$}} & \multicolumn{2}{c}{\textbf{GMC}} & \multicolumn{2}{c}{\textbf{DMC$_{\rm Att}$}} & \multicolumn{2}{c}{\textbf{GMC}} & \multicolumn{2}{c}{\textbf{DMC$_{\rm Att}$}} \\ \hline
\multicolumn{2}{c}{\textbf{90\%}}                           & \multicolumn{2}{c}{10.00}        & \multicolumn{2}{c}{10.00}        & \multicolumn{2}{c}{10.00}        & \multicolumn{2}{c}{10.00}        & \multicolumn{2}{c}{10.00}        & \multicolumn{2}{c}{10.00}        \\
\multicolumn{2}{c}{\textbf{80\%}}                           & \multicolumn{2}{c}{10.00}        & \multicolumn{2}{c}{10.00}        & \multicolumn{2}{c}{10.00}        & \multicolumn{2}{c}{10.00}        & \multicolumn{2}{c}{10.00}        & \multicolumn{2}{c}{10.00}        \\
\multicolumn{2}{c}{\textbf{70\%}}                           & \multicolumn{2}{c}{85.13}        & \multicolumn{2}{c}{62.85}        & \multicolumn{2}{c}{87.46}        & \multicolumn{2}{c}{57.62}        & \multicolumn{2}{c}{74.03}        & \multicolumn{2}{c}{60.41}        \\
\multicolumn{2}{c}{\textbf{60\%}}                           & \multicolumn{2}{c}{86.13}        & \multicolumn{2}{c}{100.00}       & \multicolumn{2}{c}{88.18}        & \multicolumn{2}{c}{99.52}        & \multicolumn{2}{c}{88.77}        & \multicolumn{2}{c}{95.71}        \\
\multicolumn{2}{c}{\textbf{50\%}}                           & \multicolumn{2}{c}{86.13}        & \multicolumn{2}{c}{100.00}       & \multicolumn{2}{c}{88.07}        & \multicolumn{2}{c}{98.57}        & \multicolumn{2}{c}{88.78}        & \multicolumn{2}{c}{99.52}        \\
\multicolumn{2}{c}{\textbf{40\%}}                           & \multicolumn{2}{c}{86.03}        & \multicolumn{2}{c}{100.00}       & \multicolumn{2}{c}{88.12}        & \multicolumn{2}{c}{99.52}        & \multicolumn{2}{c}{88.89}        & \multicolumn{2}{c}{99.52}        \\
\multicolumn{2}{c}{\textbf{30\%}}                           & \multicolumn{2}{c}{86.07}        & \multicolumn{2}{c}{100.00}       & \multicolumn{2}{c}{88.17}        & \multicolumn{2}{c}{100.00}       & \multicolumn{2}{c}{88.84}        & \multicolumn{2}{c}{100.00}       \\ \hline\hline
% \textbf{}                    & \textbf{}                    &                 &                &                 &                &                 &                &                 &                &                 &                &                 &                \\ \hline\hline
\multicolumn{14}{c}{\textbf{CIFAR-100}}                                                                                                                                                                                                                                       \\
\multicolumn{2}{c}{\multirow{2}{*}{\textbf{PruningAtt}}} & \multicolumn{4}{c}{\textbf{ResNet-18   (Threshold=10.00\%)}}        & \multicolumn{4}{c}{\textbf{ResNet-34   (Threshold=10.95\%)}}        & \multicolumn{4}{c}{\textbf{ResNet-50 (Threshold=8.80\%)}}           \\
\multicolumn{2}{c}{}                                        & \multicolumn{2}{c}{\textbf{GMC}} & \multicolumn{2}{c}{\textbf{DMC$_{\rm Att}$}} & \multicolumn{2}{c}{\textbf{GMC}} & \multicolumn{2}{c}{\textbf{DMC$_{\rm Att}$}} & \multicolumn{2}{c}{\textbf{GMC}} & \multicolumn{2}{c}{\textbf{DMC$_{\rm Att}$}} \\ \hline
\multicolumn{2}{c}{\textbf{90\%}}                           & \multicolumn{2}{c}{16.11}        & \multicolumn{2}{c}{19.04}        & \multicolumn{2}{c}{15.02}        & \multicolumn{2}{c}{10.00}        & \multicolumn{2}{c}{9.70}         & \multicolumn{2}{c}{10.00}        \\
\multicolumn{2}{c}{\textbf{80\%}}                           & \multicolumn{2}{c}{38.40}        & \multicolumn{2}{c}{20.47}        & \multicolumn{2}{c}{39.04}        & \multicolumn{2}{c}{19.04}        & \multicolumn{2}{c}{40.69}        & \multicolumn{2}{c}{26.66}        \\
\multicolumn{2}{c}{\textbf{70\%}}                           & \multicolumn{2}{c}{44.72}        & \multicolumn{2}{c}{28.57}        & \multicolumn{2}{c}{44.46}        & \multicolumn{2}{c}{25.71}        & \multicolumn{2}{c}{49.73}        & \multicolumn{2}{c}{46.66}        \\
\multicolumn{2}{c}{\textbf{60\%}}                           & \multicolumn{2}{c}{47.18}        & \multicolumn{2}{c}{61.42}        & \multicolumn{2}{c}{47.34}        & \multicolumn{2}{c}{75.71}        & \multicolumn{2}{c}{52.49}        & \multicolumn{2}{c}{86.19}        \\
\multicolumn{2}{c}{\textbf{50\%}}                           & \multicolumn{2}{c}{47.81}        & \multicolumn{2}{c}{89.04}        & \multicolumn{2}{c}{47.89}        & \multicolumn{2}{c}{98.09}        & \multicolumn{2}{c}{53.69}        & \multicolumn{2}{c}{99.04}        \\
\multicolumn{2}{c}{\textbf{40\%}}                           & \multicolumn{2}{c}{48.47}        & \multicolumn{2}{c}{100.00}       & \multicolumn{2}{c}{48.17}        & \multicolumn{2}{c}{99.52}        & \multicolumn{2}{c}{53.87}        & \multicolumn{2}{c}{100.00}       \\
\multicolumn{2}{c}{\textbf{30\%}}                           & \multicolumn{2}{c}{48.68}        & \multicolumn{2}{c}{100.00}       & \multicolumn{2}{c}{48.41}        & \multicolumn{2}{c}{100.00}       & \multicolumn{2}{c}{53.89}        & \multicolumn{2}{c}{100.00}       \\ \hline\hline
\end{tabular}}
\end{table}

\textbf{PruningAtt.} It is well known that DNNs have many layers and many parameters, leading to the possible existence of redundant parameters.
Therefore, a plagiarist can use PruningAtt to cut some redundant parameters and obtain a new model that is different from the original model but still has similar accuracy.
Using a PruningAtt algorithm, some percentage of the parameters with the smallest absolute value is set to 0. The experiments compress the parameters by 30\% to 90\% with an interval of 10\%. We then evaluate the accuracy of the model using the original test data to determine the impact on the original functionality of the model and also to assess the impact on FedRight.
Ideally, a plagiarist would like to prune the stolen model and still maintain its performance.
% Then, we use the original test data to evaluate the prediction accuracy of the pruned model.  That is to evaluate whether the pruned model can evade the copyright protection and with great value (having similar accuracy with the original one). 

Observation of the experimental results in Table \ref{jianzhi}. It can be seen that as the PruningAtt pruning percentage increases, it seriously affects the main task performance. Although the verification accuracy of the corresponding detector also gradually decreases, it still has the verification capability when the IP model still has high accuracy.

Even though some DMC of the detector has fallen below the threshold, we can see that the global model itself has also become of no value at that time. 
For example, for the three models on the CIFAR-10 dataset, when the global model is pruned by 90\%, the DMC is lowered to 10\% which is not able to determine the ownership anymore. However, the global model accuracy drops to 10\% at the same time, which is of no meaning to protect.

\begin{table}[t]
\centering
\caption{The experimental results of AdaptiveAtts where attackers have different levels of knowledge.}\label{adpt}
\resizebox{\linewidth}{!}{ \huge
\begin{tabular}{cll|crrrrrcrrrrrcrrrrr}
\hline \hline
\multicolumn{3}{c|}{\multirow{3}{*}{\textbf{Knowledge}}} & \multicolumn{6}{c}{\textbf{MNIST}}                                                                                                                                                                       & \multicolumn{6}{c}{\textbf{CIFAR-10}}                                                                                                                                                                    & \multicolumn{6}{c}{\textbf{CIFAR-100}}                                                                                                                                                                   \\
\multicolumn{3}{c|}{}                                    & \multicolumn{2}{c}{\textbf{LeNet-1}}                         & \multicolumn{2}{c}{\textbf{LeNet-4}}                                & \multicolumn{2}{c}{\textbf{LeNet-5}}                                & \multicolumn{2}{c}{\textbf{VGG-11}}                          & \multicolumn{2}{c}{\textbf{VGG-13}}                                 & \multicolumn{2}{c}{\textbf{VGG-16}}                                 & \multicolumn{2}{c}{\textbf{ResNet-18}}                       & \multicolumn{2}{c}{\textbf{ResNet-34}}                              & \multicolumn{2}{c}{\textbf{ResNet-50}}                              \\
\multicolumn{3}{c|}{}                                    & \textbf{GMC}              & \multicolumn{1}{c}{\textbf{DMC$_{\rm Att}$}} & \multicolumn{1}{c}{\textbf{GMC}} & \multicolumn{1}{c}{\textbf{DMC$_{\rm Att}$}} & \multicolumn{1}{c}{\textbf{GMC}} & \multicolumn{1}{c}{\textbf{DMC$_{\rm Att}$}} & \textbf{GMC}              & \multicolumn{1}{c}{\textbf{DMC$_{\rm Att}$}} & \multicolumn{1}{c}{\textbf{GMC}} & \multicolumn{1}{c}{\textbf{DMC$_{\rm Att}$}} & \multicolumn{1}{c}{\textbf{GMC}} & \multicolumn{1}{c}{\textbf{DMC$_{\rm Att}$}} & \textbf{GMC}              & \multicolumn{1}{c}{\textbf{DMC$_{\rm Att}$}} & \multicolumn{1}{c}{\textbf{GMC}} & \multicolumn{1}{c}{\textbf{DMC$_{\rm Att}$}} & \multicolumn{1}{c}{\textbf{GMC}} & \multicolumn{1}{c}{\textbf{DMC$_{\rm Att}$}} \\ \hline
\multicolumn{3}{c|}{\textbf{AdaptiveAtt-1}}              & \multicolumn{1}{r}{94.03} & 76.66                            & 97.77                            & 94.60                            & 97.11                            & 86.40                            & \multicolumn{1}{r}{79.80} & 62.85                            & 83.47                            & 69.52                            & 83.97                            & 66.66                            & \multicolumn{1}{r}{1.87}  & 23.80                            & 2.55                             & 20.47                            & 1.43                             & 26.19                            \\
\multicolumn{3}{c|}{\textbf{AdaptiveAtt-2}}              & \multicolumn{1}{r}{35.41} & 14.76                            & 29.39                            & 11.42                            & 37.96                            & 20.00                            & \multicolumn{1}{r}{55.23} & 33.33                            & 72.21                            & 38.09                            & 77.70                            & 42.85                            & \multicolumn{1}{r}{1.48}  & 12.38                            & 1.06                             & 7.14                             & 1.16                             & 12.85                            \\
\multicolumn{3}{c|}{\textbf{AdaptiveAtt-3}}              & \multicolumn{1}{r}{96.94} & 91.42                            & 97.99                            & 77.14                            & 98.06                            & 67.14                            & \multicolumn{1}{r}{84.85} & 70.95                            & 85.75                            & 62.85                            & 86.58                            & 81.42                            & \multicolumn{1}{r}{39.77} & 31.42                            & 39.12                            & 49.04                            & 41.63                            & 23.80                            \\ \hline \hline
\end{tabular}}
\end{table}

\textbf{CollabAtt.} We design two experimental scenarios to satisfy the inter-collaboration of malicious parties.

\textit{$CollabAtt_{up}$.} Upstream malicious client collaboration. Malicious clients participating in FL training can collaborate to eliminate model fingerprints by using local training data for adversarial retraining attacks.

\textit{$CollabAtt_{updo}$.} 
Upstream and downstream malicious parties collaborative attack. In this model, the upstream malicious client performs adversarial retraining with local training data, and then the downstream malicious party attacks the trained global model (e.g. pruning, fine-tuning), thus they implement the collaborative attack by both the upstream and downstream attackers.

We test the effect of two collaborative attacks on different proportions of malicious clients. The experimental results are shown in Table \ref{aco}. We find that FedRight remains robust in the face of collaborative attacks, the main threat to IP verification comes from downstream attackers. $CollabAtt_{up}$ cannot effectively affect the verification accuracy of the detector $M$. Facing $CollabAtt_{ftun}$ and $CollabAtt_{prun}$, the detector $M$ verification accuracy has a certain decrease, but does not affect the effective judgment of the detector $M$. 
Besides, we study how the number of malicious clients affects the FedRight, and we find that the number of malicious clients does not affect our IP statement of the global model.
Unlike the watermark embedding technique, most malicious clients can effectively remove the watermark embedded in the global model. In the FedRight framework, however, all the attacks done by the clients will only affect the performance of the global model, and the detector will be dynamically trained according to the changes in the global model to achieve the IP protection.

\begin{table}[h]
\centering
\caption{The experimental results of CollabAtts where attackers have different ways of collaboration.
}\label{aco}
\resizebox{\linewidth}{!}{ 
\begin{tabular}{cc|rrrrrr|rrrrrr}
\hline\hline
\multicolumn{2}{c|}{\multirow{3}{*}{\textbf{CollabAtt}}} & \multicolumn{6}{c|}{\textbf{40\%}}                                                                                                                                                                               & \multicolumn{6}{c}{\textbf{80\%}}                                                                                                                                                                               \\
\multicolumn{2}{c|}{}                                    & \multicolumn{2}{c}{\textbf{$CollabAtt_{up}$}}                            & \multicolumn{2}{c}{\textbf{$CollabAtt_{ftun}$}}                            & \multicolumn{2}{c|}{\textbf{$CollabAtt_{prun}$}}                            & \multicolumn{2}{c}{\textbf{$CollabAtt_{up}$}}                            & \multicolumn{2}{c}{\textbf{$CollabAtt_{ftun}$}}                            & \multicolumn{2}{c}{\textbf{$CollabAtt_{prun}$}}                            \\
\multicolumn{2}{c|}{}                                    & \multicolumn{1}{c}{\textbf{GMC}} & \multicolumn{1}{c}{\textbf{DMC$_{\rm Att}$}} & \multicolumn{1}{c}{\textbf{GMC}} & \multicolumn{1}{c}{\textbf{DMC$_{\rm Att}$}} & \multicolumn{1}{c}{\textbf{GMC}} & \multicolumn{1}{c|}{\textbf{DMC$_{\rm Att}$}} & \multicolumn{1}{c}{\textbf{GMC}} & \multicolumn{1}{c}{\textbf{DMC$_{\rm Att}$}} & \multicolumn{1}{c}{\textbf{GMC}} & \multicolumn{1}{c}{\textbf{DMC$_{\rm Att}$}} & \multicolumn{1}{c}{\textbf{GMC}} & \multicolumn{1}{c}{\textbf{DMC$_{\rm Att}$}} \\ \hline
\multirow{3}{*}{\textbf{MNIST}}     & \textbf{LeNet-1}   & 96.78                            & 100.00                           & 96.93                            & 98.57                            & 94.07                            & 90.95                             & 97.18                            & 100.00                           & 96.85                            & 81.90                            & 87.16                            & 71.42                            \\
                                    & \textbf{LeNet-4}   & 98.24                            & 100.00                           & 97.89                            & 95.71                            & 97.32                            & 94.28                             & 98.68                            & 100.00                           & 98.32                            & 93.33                            & 98.05                            & 94.28                            \\
                                    & \textbf{LeNet-5}   & 98.52                            & 100.00                           & 98.47                            & 97.62                            & 97.89                            & 88.57                             & 98.65                            & 100.00                           & 98.10                            & 96.67                            & 97.34                            & 82.85                            \\ \hline
\multirow{3}{*}{\textbf{CIFAR-10}}  & \textbf{VGG-11}    & 86.37                            & 100.00                           & 85.84                            & 98.57                            & 85.91                            & 100.00                            & 85.24                            & 100.00                           & 85.54                            & 96.67                            & 84.89                            & 100.00                           \\
                                    & \textbf{VGG-13}    & 87.69                            & 100.00                           & 88.12                            & 94.29                            & 87.97                            & 99.04                             & 87.85                            & 100.00                           & 87.31                            & 85.71                            & 87.26                            & 100.00                           \\
                                    & \textbf{VGG-16}    & 88.01                            & 100.00                           & 88.83                            & 99.52                            & 87.99                            & 100.00                            & 88.92                            & 100.00                           & 88.29                            & 97.62                            & 88.51                            & 100.00                           \\ \hline
\multirow{3}{*}{\textbf{CIFAR-100}} & \textbf{ResNet-18} & 40.32                            & 100.00                           & 45.49                            & 61.90                            & 37.12                            & 87.14                             & 13.78                            & 100.00                           & 45.06                            & 78.10                            & 12.75                            & 83.80                            \\
                                    & \textbf{ResNet-34} & 43.36                            & 100.00                           & 44.31                            & 89.05                            & 42.29                            & 83.80                             & 13.05                            & 100.00                           & 46.04                            & 52.38                            & 12.54                            & 97.61                            \\
                                    & \textbf{ResNet-50} & 45.49                            & 100.00                           & 49.96                            & 65.90                            & 41.53                            & 96.19                             & 11.26                            & 100.00                           & 50.52                            & 64.29                            & 8.08                             & 90.95                            \\ \hline\hline
\end{tabular}}
\end{table}

\textbf{AdaptiveAtt.} We designed AdaptiveAtt with different levels of adversary capability to test the robustness of IP verification.

\textit{AdaptiveAtt-1.} 
In the FL scenario, it is difficult to obtain all the training data of the client, we assume that the attacker has has 10\% of the original training samples.
They can generate adversarial examples using the original training samples, and perform adversarial retraining on the original model to eliminate fingerprints.

\textit{AdaptiveAtt-2.} The attacker has the all key samples. Adversarial examples are generated using key samples. Then adversarial retraining of the original model is conducted to eliminate fingerprints. 

\textit{AdaptiveAtt-3.} 
The attacker has AdaptiveAtt-3 containing AdaptiveAtt-1 and AdaptiveAtt-2,
which first eliminates the fingerprints by using adversarial retraining of the key sample and then retraining with the original training sample.

We tested on nine models, and the experimental results are shown in Table \ref{adpt}, where we found that the attacker using AdaptiveAtt-1 cannot remove the model fingerprint verification.
And while using AdaptiveAtt-2 is effective in removing model fingerprints, it has a greater impact on the prediction performance of the main task, which renders the attack invalid since the main task performance is severely impaired.
Using AdaptiveAtt-3, the model accuracy is improved by retraining the original training samples after the adversarial retraining using AdaptiveAtt-2. However the model fingerprint can still be validated effectively.

\textbf{CopEvaAtt.} To evade the legitimate owner's verification, the attacker can try to construct a detector to detect whether the queried sample is a clean sample or possibly a fingerprinting sample. Once the detector determines that the queried instance is a fingerprinting sample, the attacker can diliberately return a random label.

However, the data of each client is kept locally and not shared during the federation learning process. It is also not sure that the data are independent and identically distributed (IID) among clients.
It is undesirable for malicious clients to try to defend against copyright verification by CopEvaAtt.
This detection method is more costly for the attacker and difficult to deploy in practice.
The success rate of defending against copyright verification will come at the cost of more false positives, which reduces the utility of the attacker model for other users.

\begin{center}
\fcolorbox{black}{white!20}{\parbox{0.97\linewidth}
{
\emph{\textbf{Answer to RQ3:}}
FedRight shows strong robustness under model modification attacks including both FtuningAtt and PruningAtt. It maintains its validity in the face of AdaptiveAtts, i.e., it either maintains high verification effectiveness or invalids the purpose of stealing a model by lowering the accuracy to an undesired level.
}
}
\end{center}

\subsection{RQ4: Comparison with Watermarking Technology}
In this section, we compare FedRight with the watermarking IP protection. The static generation of the adversarial example 
FedRight$_{\rm S}$ is also added to the comparison experiment to illustrate the advantages of the dynamic generation of the adversarial example FedRight$_{\rm D}$. 

\subsubsection{Impact on Main Performance}

\begin{figure}[h]
\centering{
        \includegraphics[width=0.6\textwidth]{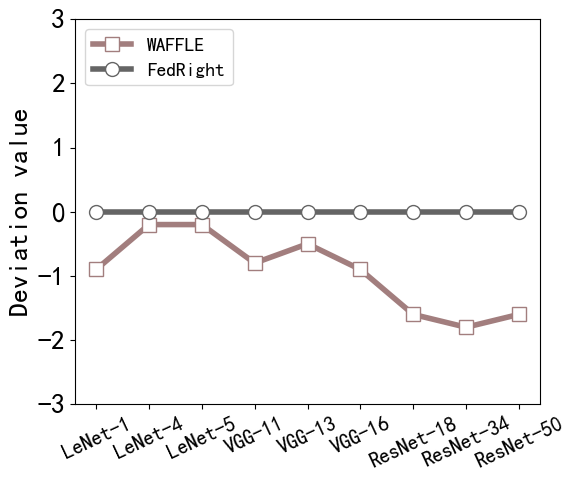} }

\caption{
The accuracy of the global models protected by FedRight and WAFFLE. It is measured by the deviation from the baseline which is the global model trained without any copyright protection.
}\label{deviation}
\end{figure}

RQ1 answers that FedRight has a natural advantage in terms of fidelity. To better confirm its fidelity, we performed a comparative evaluation of the watermark embedding techniques-WAFFLE and FedRight. The results are shown in Fig.\ref{deviation}, FedRight shows the same level of global model accuracy as the normal FL model, while the accuracy of the watermarked FL model decreases by 0.94 \% on average.
The reason is that our FedRight IP protection is independent of the federated training process and does not have any impact on FL. The embedding of watermark changes the training settings of FL and needs to embed the watermark image into the global model.
Although with the development of technology, the impact from watermarking technology on the main task performance is gradually reduced, it is still difficult to eliminate this inherent defect.

\subsubsection{Time Overhead}

\begin{table}[h]
\centering
\caption{The time overhead added to the main task FL training by WAFFLE and FedRight.}
\resizebox{\textwidth}{!}{
\begin{tabular}{cl|crrcrrcrr}
\hline \hline
\multicolumn{2}{c|}{\multirow{3}{*}{~~~~\textbf{Method}~~~~}} & \multicolumn{9}{c}{\textbf{Time(second)}}                                                                                                                                                                                                                                                                                          \\
\multicolumn{2}{c|}{}                                 & \multicolumn{3}{c}{\textbf{epoch = 80}}                                                                  & \multicolumn{3}{c}{\textbf{epoch = 100}}                                                                 & \multicolumn{3}{c}{\textbf{epoch = 100}}                                                                      \\
\multicolumn{2}{c|}{}                                 & \textbf{LeNet-1}           & \multicolumn{1}{c}{\textbf{LeNet-4}} & \multicolumn{1}{c}{\textbf{LeNet-5}} & \textbf{VGG-11}             & \multicolumn{1}{c}{\textbf{VGG-13}} & \multicolumn{1}{c}{\textbf{VGG-16}} & \textbf{ResNet-18}          & \multicolumn{1}{c}{\textbf{ResNet-34}} & \multicolumn{1}{c}{\textbf{ResNet-50}} \\ \hline
\multicolumn{2}{c|}{\textbf{WAFFLE}}                  & \multicolumn{1}{r}{424.46} & 462.75                               & 480.50                               & \multicolumn{1}{r}{1645.37} & 1825.40                             & 2185.46                             & \multicolumn{1}{r}{2415.42} & 2909.91                                & 5500.70                                \\
\multicolumn{2}{c|}{\textbf{FedRight}}                & \multicolumn{1}{r}{\textbf{0.00}}   & \textbf{0.00}                                & \textbf{0.00}                               & \multicolumn{1}{r}{\textbf{0.00}}    & \textbf{0.00}                               & \textbf{0.00}                              & \multicolumn{1}{r}{\textbf{0.00}}    & \textbf{0.00}                                  & \textbf{0.00}                                   \\ \hline \hline
\end{tabular}}\label{times}
\end{table}

\begin{figure}[h]
\centering{
        \includegraphics[width=0.7\textwidth]{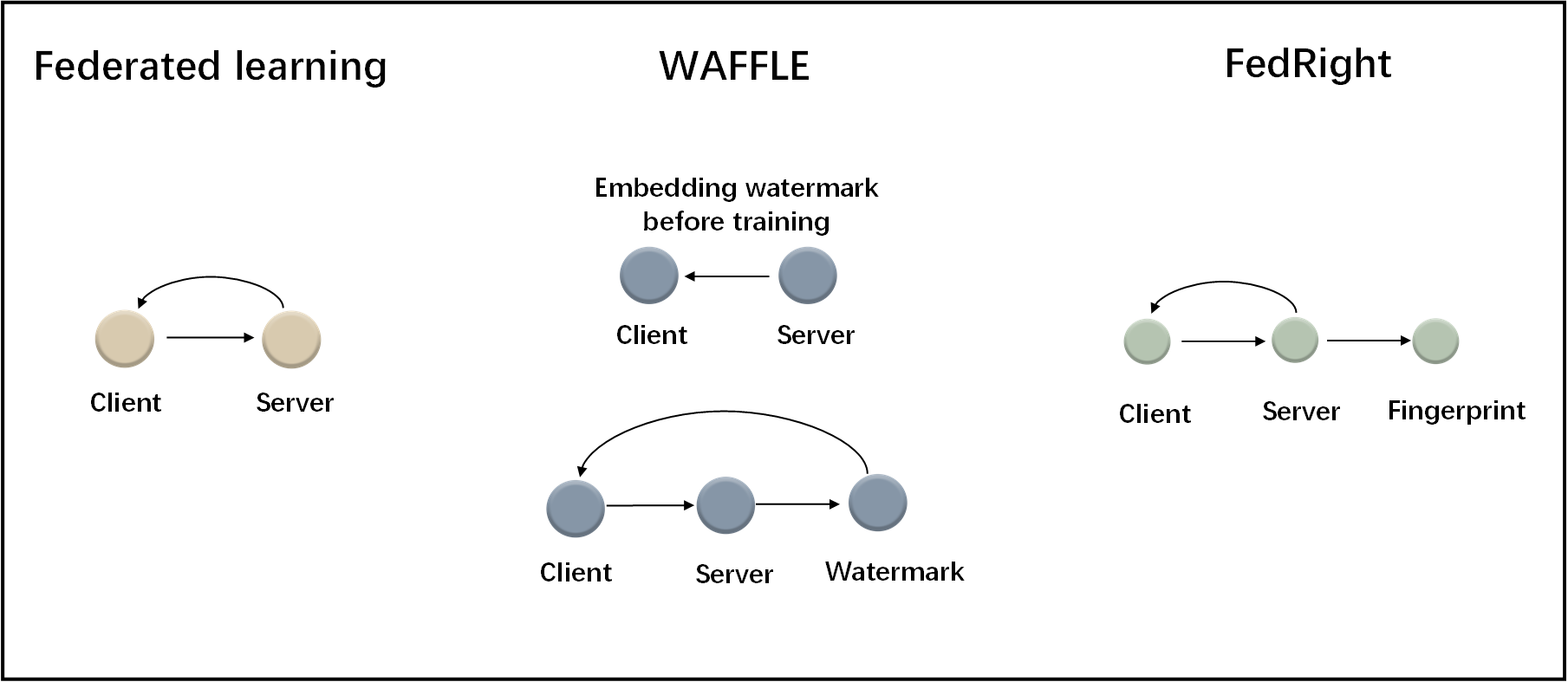} }

\caption{
Workflow illustrations of the vanilla FL, WAFFLE and FedRight.
}\label{time}
\end{figure}

We compare the time overhead caused by the WAFFLE and FedRight.
As shown in Fig.\ref{time},
it can be observed intuitively that WAFFLE has changed the settings of the federated training. The server needs to embed the watermark after each epoch of aggregation operation, and then distribute the global model to each client. 
WAFFLE is at the expense of the main task training performance and efficiency.
Differently, FedRight does not have any impact on the main task training, and the model fingerprinting is independent of the FL training. 
% So the process of issuing the global model by the server is not blocked.
In Table \ref{times} we give the detailed time overhead added by the two copyright protection mechanisms for federated training. WAFFLE adds an average of 1983.89 seconds of time overhead for all models.
The time added on training a large dataset such as CIFAR100 is even more significant, with an average increase of 3608.68 seconds.

\begin{figure}[h]
\centering
    {
    \subfigure[LeNet-5]{
    \includegraphics[width=0.3\linewidth]{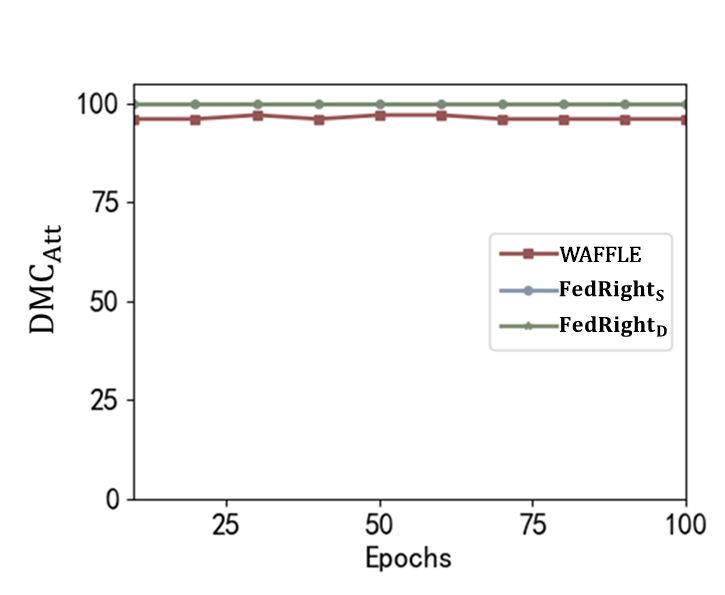}}
    \hspace{0.01\linewidth}
    \subfigure[VGG-11]{
    \includegraphics[width=0.3\linewidth]{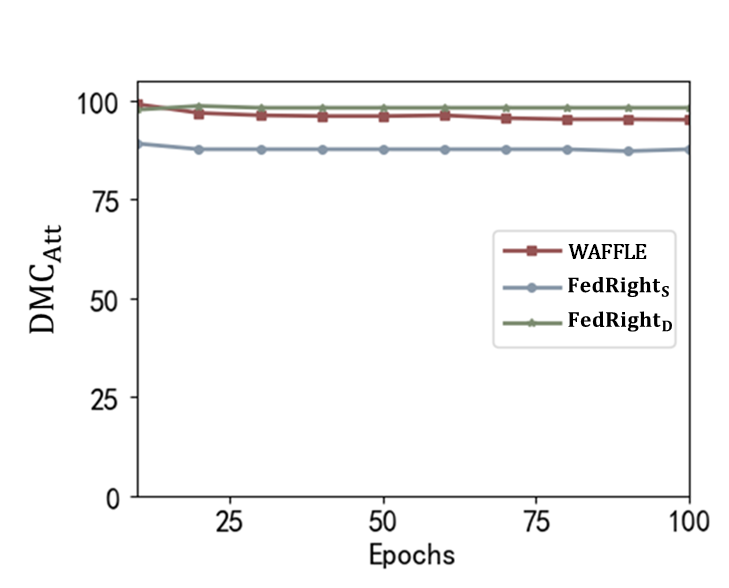}}
    \hspace{0.01\linewidth}
    \subfigure[ResNet-18]{
    \includegraphics[width=0.3\linewidth]{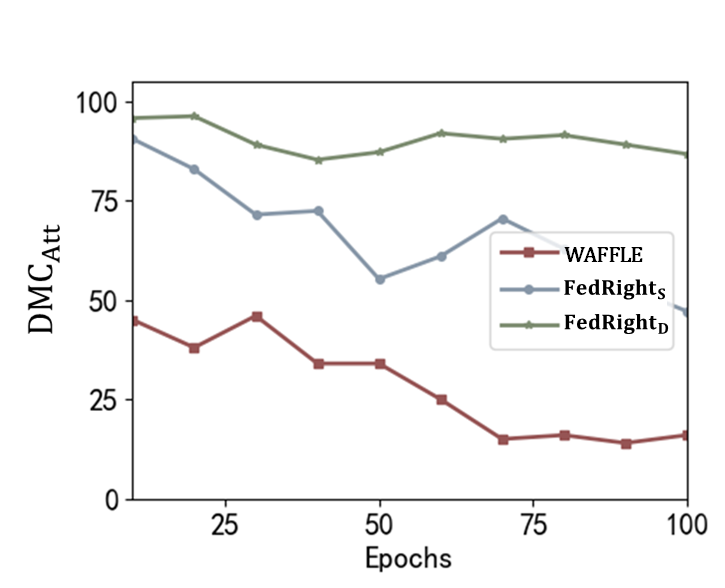}}} 
\caption{The accuracy of IP verification changes during PruningAtt for 100 epochs using 10\% of the training set.
}\label{2}
\end{figure}

% \usepackage{multirow}
\begin{table}[h]
\centering
\caption{
Comparison of the robustness of WAFFLE and FedRight against PruningAtt.
}\label{3}
\resizebox{1\textwidth}{!}{\Huge
\begin{tabular}{clrrrrrrrrrrrrrrrrrrrr}
\hline \hline
\multicolumn{2}{c}{\multirow{3}{*}{\textbf{PruningAtt}}} & \multicolumn{6}{c}{\textbf{LeNet-5}}                                                                                                                                                                            & \multicolumn{1}{c}{\textbf{}} & \multicolumn{6}{c}{\textbf{VGG-11}}                                                                                                                                                                              & \multicolumn{1}{c}{\textbf{}} & \multicolumn{6}{c}{\textbf{ResNet-18}}                                                                                                                                                                          \\
\multicolumn{2}{c}{}                                        & \multicolumn{2}{c}{\textbf{WAFFLE}}                                 & \multicolumn{2}{c}{\textbf{FedRight$_{\rm S}$}}                                 & \multicolumn{2}{c}{\textbf{FedRight$_{\rm D}$}}                                 & \multicolumn{1}{c}{\textbf{}} & \multicolumn{2}{c}{\textbf{WAFFLE}}                                 & \multicolumn{2}{c}{\textbf{FedRight$_{\rm S}$}}                                 & \multicolumn{2}{c}{\textbf{FedRight$_{\rm D}$}}                                 & \multicolumn{1}{c}{\textbf{}} & \multicolumn{2}{c}{\textbf{WAFFLE}}                                 & \multicolumn{2}{c}{\textbf{FedRight$_{\rm S}$}}                                 & \multicolumn{2}{c}{\textbf{FedRight$_{\rm D}$}}                                 \\
\multicolumn{2}{c}{}                                        & \multicolumn{1}{c}{\textbf{GMC}} & \multicolumn{1}{c}{\textbf{DMC$_{\rm Att}$}} & \multicolumn{1}{c}{\textbf{GMC}} & \multicolumn{1}{c}{\textbf{DMC$_{\rm Att}$}} & \multicolumn{1}{c}{\textbf{GMC}} & \multicolumn{1}{c}{\textbf{DMC$_{\rm Att}$}} & \multicolumn{1}{c}{\textbf{}} & \multicolumn{1}{c}{\textbf{GMC}} & \multicolumn{1}{c}{\textbf{DMC$_{\rm Att}$}} & \multicolumn{1}{c}{\textbf{GMC}} & \multicolumn{1}{c}{\textbf{DMC$_{\rm Att}$}} & \multicolumn{1}{c}{\textbf{GMC}} & \multicolumn{1}{c}{\textbf{DMC$_{\rm Att}$}} & \multicolumn{1}{c}{\textbf{}} & \multicolumn{1}{c}{\textbf{GMC}} & \multicolumn{1}{c}{\textbf{DMC}} & \multicolumn{1}{c}{\textbf{GMC}} & \multicolumn{1}{c}{\textbf{DMC$_{\rm Att}$}} & \multicolumn{1}{c}{\textbf{GMC}} & \multicolumn{1}{c}{\textbf{DMC$_{\rm Att}$}} \\ \hline
\multicolumn{2}{c}{\textbf{90\%}}                             & 72.42                            & 44.00                            & 60.74                            & 35.71                            & 72.39                            & \textbf{46.60}                   &                               & 10.00                            & 10.00                            & 10.00                            & 10.00                            & 10.00                            & \textbf{10.00}                   &                               & 14.34                            & 0.00                             & 15.78                            & 10.00                            & 16.11                            & \textbf{19.04}                   \\
\multicolumn{2}{c}{\textbf{80\%}}                             & 90.72                            & 68.00                            & 89.63                            & 67.33                            & 90.79                            & \textbf{69.60}                   &                               & 10.00                            & 10.00                            & 10.00                            & 10.00                            & 10.00                            & \textbf{10.00}                   &                               & 38.20                            & 1.00                             & 10.00                            & 10.00                            & 38.40                            & \textbf{20.47}                   \\
\multicolumn{2}{c}{\textbf{70\%}}                             & 90.64                            & 87.00                            & 92.43                            & 86.35                            & 92.36                            & \textbf{87.40}                   &                               & 83.62                            & 43.80                            & 84.97                            & 59.80                            & 85.13                            & \textbf{62.85}                   &                               & 44.52                            & 11.24                            & 45.55                            & 14.28                            & 44.72                            & \textbf{28.57}                   \\
\multicolumn{2}{c}{\textbf{60\%}}                             & 96.21                            & 100.00                           & 97.66                            & 100.00                           & 96.65                            & \textbf{100.00}                  &                               & 84.52                            & 62.04                            & 85.52                            & 99.04                            & 86.13                            & \textbf{100.00}                  &                               & 45.73                            & 30.46                            & 47.04                            & 31.42                            & 47.18                            & \textbf{61.42}                   \\
\multicolumn{2}{c}{\textbf{50\%}}                             & 96.89                            & 100.00                           & 98.15                            & 100.00                           & 96.84                            & \textbf{100.00}                  &                               & 85.12                            & 78.83                            & 85.48                            & 99.52                            & 86.13                            & \textbf{100.00}                  &                               & 46.24                            & 67.35                            & 47.25                            & 86.95                            & 47.81                            & \textbf{89.04}                   \\
\multicolumn{2}{c}{\textbf{40\%}}                             & 98.10                            & 100.00                           & 98.49                            & 100.00                           & 98.01                            & \textbf{100.00}                  &                               & 85.25                            & 92.51                            & 85.47                            & 100.00                           & 86.03                            & \textbf{100.00}                  &                               & 46.63                            & 93.18                            & 47.76                            & 99.52                            & 48.47                            & \textbf{100.00}                  \\
\multicolumn{2}{c}{\textbf{30\%}}                             & 98.25                            & 100.00                           & 98.46                            & 100.00                           & 98.46                            & \textbf{100.00}                  &                               & 85.31                            & 100.00                           & 85.51                            & 100.00                           & 86.07                            & \textbf{100.00}                  &                               & 47.35                            & 100.00                           & 48.71                            & 100.00                           & 48.68                            & \textbf{100.00}                  \\ \hline \hline
\end{tabular}}
\end{table}

\subsubsection{Ability to Resist Attacks}
We consider two real-world scenarios in which attackers might execute the attack methods. (1) MaliClientAtt: upstream attackers perform the removal attack to interfere the copyright protection implementation during training. (2) Downstream attackers use FtuningAtt and PruningAtt to clean the global model.

% \usepackage{multirow}
\begin{table}[]
\centering
\caption{The impact of 40\% and 80\% malicious clients performing removal attack during training.}\label{1}
\resizebox{0.8\linewidth}{!}{ 
\begin{tabular}{cc|rrrrrr}
\hline \hline
\multicolumn{2}{c|}{\multirow{2}{*}{\textbf{Model}}} & \multicolumn{3}{c}{\textbf{40\%}}                                                                              & \multicolumn{3}{c}{\textbf{80\%}}                                                                              \\
\multicolumn{2}{l|}{}                                & \multicolumn{1}{c}{\textbf{WAFFLE}} & \multicolumn{1}{c}{\textbf{FedRight$_{\rm S}$}} & \multicolumn{1}{c}{\textbf{FedRight$_{\rm D}$}} & \multicolumn{1}{c}{\textbf{WAFFLE}} & \multicolumn{1}{c}{\textbf{FedRight$_{\rm S}$}} & \multicolumn{1}{c}{\textbf{FedRight$_{\rm D}$}} \\ \hline
\multirow{9}{*}{\textbf{DMC$_{\rm Att}$}}  & \textbf{LeNet-1}    & 25.31                              & \textbf{100.00}                     & \textbf{100.00}                     & 14.31                              & \textbf{100.00}                     & \textbf{100.00}                     \\
                               & \textbf{LeNet-4}    & 24.68                              & \textbf{100.00}                     & \textbf{100.00}                     & 12.68                              & \textbf{100.00}                     & \textbf{100.00}                     \\
                               & \textbf{LeNet-5}    & 25.42                              & \textbf{100.00}                     & \textbf{100.00}                     & 12.42                              & \textbf{100.00}                     & \textbf{100.00}                     \\
                               & \textbf{VGG-11}     & 20.41                              & \textbf{100.00}                     & \textbf{100.00}                     & 10.68                              & \textbf{100.00}                     & \textbf{100.00}                     \\
                               & \textbf{VGG-13}     & 19.04                              & \textbf{100.00}                     & \textbf{100.00}                     & 9.54                               & \textbf{100.00}                     & \textbf{100.00}                     \\
                               & \textbf{VGG-16}     & 21.52                              & \textbf{100.00}                     & \textbf{100.00}                     & 11.52                              & \textbf{100.00}                     & \textbf{100.00}                     \\
                               & \textbf{ResNet-18}  & 0.00                               & \textbf{100.00}                     & \textbf{100.00}                     & 0.00                               & \textbf{100.00}                     & \textbf{100.00}                     \\
                               & \textbf{ResNet-34}  & 0.00                               & \textbf{100.00}                     & \textbf{100.00}                     & 0.00                               & \textbf{100.00}                     & \textbf{100.00}                     \\
                               & \textbf{ResNet-50}  & 0.00                               & \textbf{100.00}                     & \textbf{100.00}                     & 0.00                               & \textbf{100.00}                     & \textbf{100.00}                     \\ \hline \hline
\end{tabular}}
\end{table}

MaliClientAtt: In the federation training process, we set up 40\% and 80\% malicious clients to perform watermark removal attack method on the global model, respectively.
The experimental results are shown in Table \ref{1}.
Facing the removal attack, the overall watermark verification accuracy of WAFFLE has been lowered to less than 30\%.  
Even worse, the verification has been failed on the CIFAR-100-based model.
The reason may be that for the main task with more data classes, the difficulty of embedding watermarks may increase, leading to easier attacks.

At present, watermark removal attacks and watermark overlay attacks are becoming more and more advanced. 
FedRight protects IP based on model fingerprints. It does not introduce watermarks to model internal parameters, so it effectively avoids various  attacks against wartermarking.

We then compare the robustness of WAFFLE and FedRight against FtuningAtt and PruningAtt.
The experimental results of FtuningAtt are shown in Fig.\ref{2}, and the experimental results of PruningAtt are shown in Table \ref{3}.
Compared with the watermark embedding technique WAFFLE, FedRight$_{\rm D}$ shows its own advantages,
In the face of FtuningAtt and PruningAtt, the overall average verification accuracy of FedRight$_{\rm D}$ is higher than that of WAFFLE by 12.64\% and 9.79\%, respectively.
We also compare FedRight$_{\rm S}$ with FedRight$_{\rm D}$,
because the parameters of the global model are updated in each training epoch, the FedRight$_{\rm S}$ may cause the fingerprints generated in the previous epoch to be invalid in the next epoch without adaptive capability. The FedRight$_{\rm D}$, on the other hand, effectively solves this problem. The validity of the model fingerprint is checked using Algorithm~\ref{en} so that the model features of the current epoch can be added adaptively to achieve an enhanced fingerprint. 
This makes the overall verification accuracy of FedRight$_{\rm D}$ higher than FedRight$_{\rm S}$ by 11.32\% and 4.05\% on average, respectively, in the face of FtuningAtt and PruningAtt.

\begin{center}
\fcolorbox{black}{white!20}{\parbox{0.97\linewidth}
{
\emph{\textbf{Answer to RQ4:}}
The FedRight technology based on model fingerprints effectively makes up for the inherent defects of watermarking technology and improves the performance by 11.22\%. The dynamically enhanced FedRight$_{\rm D}$ effectively solves the fingerprint invalidation problem and improves the performance by 7.69\% on average compared to FedRight$_{\rm S}$.
}
}
\end{center}

\subsection{RQ5: Parameter Sensitivity}
In this section, we conduct experiments to test whether
the number of clients participating in training,
the non-independent and identically distributed (Non-IID) data distribution and hyperparameter settings have an impact on the effectiveness of FedRight.

\subsubsection{Number of Clients}

To verify the impact of the number of clients on the detector  sensitivity. we test the effectiveness of the detector at K $\in$ $\left \{  5,20,100\right \}$ and also conduct the detector's robustness against adaptive attacks. The experimental results are shown in Tables \ref{k} and \ref{kr}. 
From Table \ref{k}, it can be intuitively found that the different number of clients in FL training mainly affects the global model accuracy GMC, and has no effect on the detector verification accuracy DMC. In Table \ref{kr}, even in the training scenario with client K=100, AdaptiveAtt does not lead to the failure of the detector verification. The reason is that the number of clients mainly affects the global model, while the detector will train according to the changes in the global model as a way to achieve IP protection. 
Therefore, when FL training involves a large number of clients, the effectiveness of the detector can still be ensured.

% \usepackage{multirow}

\begin{table}[h]
\centering
\caption{The impact of the number of clients $K$ on FedRight.}\label{k}
\resizebox{0.7\linewidth}{!}{ \tiny
\begin{tabular}{ll|rrr}
\hline \hline
\multicolumn{2}{c}{}                                                    & \multicolumn{1}{c}{~~~\textbf{LeNet-5}~~} & \multicolumn{1}{c}{~~~\textbf{VGG-11}~~} & \multicolumn{1}{c}{~~~\textbf{ResNet-18}~~} \\ \hline
\multirow{2}{*}{\textbf{$K$=5}\quad}  & \multicolumn{1}{c|}{\textbf{\quad GMC\quad\quad}} & 98.12                                & 85.86                               & 51.26                                  \\
                                    & \multicolumn{1}{c|}{\textbf{DMC}} & 100.00                               & 100.00                              & 100.00                                 \\ \hline
\multirow{2}{*}{\textbf{$K$=20}} &  \multicolumn{1}{c|}{\textbf{GMC}} & 98.99                                & 87.83                               & 51.71                                  \\
                                    & \multicolumn{1}{c|}{\textbf{DMC}} & 100.00                               & 100.00                              & 100.00                                 \\ \hline
\multirow{2}{*}{\textbf{$K$=100}} & \multicolumn{1}{c|}{\textbf{GMC}} & 98.57                                & 87.31                               & 56.60                                  \\
                                    & \multicolumn{1}{c|}{\textbf{DMC}} & 100.00                               & 100.00                              & 100.00                                 \\ \hline \hline
\end{tabular}}
\end{table}

\begin{table}[h]
\centering
\caption{The impact of the number of clients $K$ on the robustness of FedRight.}\label{kr}
\resizebox{\linewidth}{!}{ \huge
\begin{tabular}{cll|crrrrrcrrrrrcrrrrr}
\hline \hline
\multicolumn{3}{c|}{\multirow{3}{*}{\textbf{Knowledge}}} & \multicolumn{6}{c}{\textbf{$K$=5}}                                                                                                                                                                         & \multicolumn{6}{c}{\textbf{$K$=20}}                                                                                                                                                                        & \multicolumn{6}{c}{\textbf{$K$=100}}                                                                                                                                                                       \\
\multicolumn{3}{c|}{}                                    & \multicolumn{2}{c}{\textbf{LeNet-5}}                         & \multicolumn{2}{c}{\textbf{VGG-11}}                                 & \multicolumn{2}{c}{\textbf{ResNet-18}}                              & \multicolumn{2}{c}{\textbf{LeNet-5}}                         & \multicolumn{2}{c}{\textbf{VGG-11}}                                 & \multicolumn{2}{c}{\textbf{ResNet-18}}                              & \multicolumn{2}{c}{\textbf{LeNet-5}}                         & \multicolumn{2}{c}{\textbf{VGG-11}}                                 & \multicolumn{2}{c}{\textbf{ResNet-18}}                              \\
\multicolumn{3}{c|}{}                                    & \textbf{GMC}              & \multicolumn{1}{c}{\textbf{DMC$_{\rm Att}$}} & \multicolumn{1}{c}{\textbf{GMC}} & \multicolumn{1}{c}{\textbf{DMC$_{\rm Att}$}} & \multicolumn{1}{c}{\textbf{GMC}} & \multicolumn{1}{c}{\textbf{DMC$_{\rm Att}$}} & \textbf{GMC}              & \multicolumn{1}{c}{\textbf{DMC$_{\rm Att}$}} & \multicolumn{1}{c}{\textbf{GMC}} & \multicolumn{1}{c}{\textbf{DMC$_{\rm Att}$}} & \multicolumn{1}{c}{\textbf{GMC}} & \multicolumn{1}{c}{\textbf{DMC$_{\rm Att}$}} & \textbf{GMC}              & \multicolumn{1}{c}{\textbf{DMC$_{\rm Att}$}} & \multicolumn{1}{c}{\textbf{GMC}} & \multicolumn{1}{c}{\textbf{DMC$_{\rm Att}$}} & \multicolumn{1}{c}{\textbf{GMC}} & \multicolumn{1}{c}{\textbf{DMC$_{\rm Att}$}} \\ \hline
\multicolumn{3}{c|}{\textbf{AdaptiveAtt-1}}              & \multicolumn{1}{r}{97.11} & 86.40                            & 79.80                            & 62.85                            & 1.87                             & 23.80                            & \multicolumn{1}{r}{98.31} & 99.52                            & 84.43                            & 86.66                            & 2.10                             & 28.57                            & \multicolumn{1}{r}{93.13} & 65.71                            & 84.22                             & 62.85                             & 1.72                             & 43.33                            \\
\multicolumn{3}{c|}{\textbf{AdaptiveAtt-2}}              & \multicolumn{1}{r}{37.96} & 20.00                            & 55.23                            & 33.33                            & 1.48                             & 12.38                            & \multicolumn{1}{r}{47.65} & 12.38                            & 62.84                            & 57.14                            & 0.97                             & 32.85                            & \multicolumn{1}{r}{23.20} & 5.23                             & 60.73                                & 31.90                                & 1.18                             & 30.00                            \\
\multicolumn{3}{c|}{\textbf{AdaptiveAtt-3}}              & \multicolumn{1}{r}{98.06} & 67.14                            & 84.85                            & 70.95                            & 39.77                            & 31.42                            & \multicolumn{1}{r}{98.36} & 56.19                            & 86.71                            & 74.28                            & 42.76                            & 49.04                            & \multicolumn{1}{r}{97.22} & 60.00                            & 87.41                                & 87.61                                & 36.30                            & 26.66                            \\ \hline \hline
\end{tabular}}
\end{table}

\subsubsection{Data Distribution Sensitivity}
In real-world scenarios,
the data among clients is heterogeneous,
we consider using Dirichlet distribution to divide the training data among clients.
Specifically, we sample $Dir(\beta)$
and divide the dataset according to the distribution of concentration parameter $\beta$, and then assign them to each client.
$Dir(\beta)$ is the Dirichlet distribution with $\beta$.
With the above partitioning strategy,
the gap between the number of categories and samples owned by the clients grows as the $\beta$ value decreases.
We use $\beta$=0.1, $\beta$=0.5 and $\beta$=1 in the experiments to explore the sensitivity of data distribution to property rights protection.

% \usepackage{multirow}
\begin{table}[h]
\centering
\caption{The influence of the data distribution parameter $\beta$ on FedRight.}\label{c1}
\resizebox{0.7\linewidth}{!}{ \tiny
\begin{tabular}{ll|rrr}
\hline \hline
\textbf{}                       & \textbf{}    & \multicolumn{1}{c}{\textbf{~~~~LeNet-5~~}} & \multicolumn{1}{c}{\textbf{~~~~VGG-11~~}} & \multicolumn{1}{c}{\textbf{~~~~ResNet-18~~}} \\ \hline
\multirow{2}{*}{ \textbf{$\beta$=0.1} \quad} & \textbf{GMC\quad} & 94.01                                & 77.29                               & 38.72                                  \\
                                & \textbf{DMC} & 100.00                               & 100.00                              & 100.00                                 \\ \hline
\multirow{2}{*}{~\textbf{$\beta$=0.5}} & \textbf{GMC} & 96.18                                & 83.33                               & 48.92                                  \\
                                & \textbf{DMC} & 100.00                               & 100.00                              & 100.00                                 \\ \hline
\multirow{2}{*}{~\textbf{$\beta$=1.0}}   & \textbf{GMC} & 98.12                                & 85.86                               & 51.26                                  \\
                                & \textbf{DMC} & 100.00                               & 100.00                              & 100.00                                 \\ \hline \hline
\end{tabular}}
\end{table}

From the experimental results of Table~\ref{c1}, it is found that FedRight is not affected by differences in data distribution. The reason is that in the process of generating the model fingerprint, the fingerprint will change with the change of the global model.
However, the accuracy of global GMC is decreased as the value of $\beta$ decreases. This is the widely acknowledged issue faced by FL where the data among clients is Non-IID distribution.

\subsubsection{Key Samples}

\begin{table}[h]
\centering
\caption{The influence of the key samples on FedRight.}\label{keyn}
\resizebox{0.7\linewidth}{!}{ \tiny
\begin{tabular}{ccc|rrr}
\hline\hline
\textbf{}           & \textbf{}          & \textbf{}    & \multicolumn{1}{c}{\textbf{LeNet-5}} & \multicolumn{1}{c}{\textbf{VGG-11}} & \multicolumn{1}{c}{\textbf{ResNet-18}} \\ \hline
\multicolumn{3}{c|}{\textbf{GMC}}                       & 98.12                                & 85.86                               & 51.26                                  \\ \hline
\multicolumn{2}{c}{\textbf{Scenic samples}} & \textbf{DMC} & 100.00                               & 100.00                              & 100.00                                 \\ \hline
\multicolumn{2}{c}{\textbf{Face samples}} & \textbf{DMC} & 100.00                               & 100.00                              & 100.00                                 \\ \hline
\multicolumn{2}{c}{\textbf{Traffic instruction samples}} & \textbf{DMC} & 100.00                               & 100.00                              & 100.00                                 \\ \hline\hline
\end{tabular}}
\end{table}

We collect three key samples different from the federated training data for generating model fingerprints. The experimental results are shown in Table \ref{keyn}. The FedRight is not sensitive to the key samples, and the detector $M$ can all achieve 100\% validity. The reason is that the key samples are only part of the carriers of the model fingerprints.
The model fingerprint is formed by extracting the unique features of the model and adding them to the carrier, in which the unique features are more important than the carrier.
Therefore, the type of key sample does not affect the validity of the model fingerprint.

\subsubsection{Adversarial Example Generation Methods}

This experiment is to explore whether other adversarial example generation methods can also be used in FedRight.
As shown in Table.\ref{ad}, when using FGSM, C\&W, and PGD targeted attacks to generate the model fingerprints, FedRight can also achieve good performance. 
% This shows that the Fedright framework has universality.

\begin{table}[t]
\centering
\caption{Different adversarial example generation methods for generating model fingerprints.}
\resizebox{0.7\linewidth}{!}{ \tiny
\begin{tabular}{ccccc}
\hline \hline
             & \textbf{}                          & \textbf{LeNet-5\quad} & \textbf{VGG-11\quad} & \textbf{ResNet-18\quad} \\ \hline
\textbf{}    & \multicolumn{1}{c|}{\textbf{C\&W\quad}}  & 100              & 100             & 100                \\
\textbf{DMC\quad} & \multicolumn{1}{c|}{\textbf{PGD}}   & 100              & 100             & 100                \\
\textbf{}    & \multicolumn{1}{c|}{\textbf{FGSM}} & 100              & 100             & 100                \\ \hline \hline
\end{tabular}}\label{ad}
\end{table}

\subsubsection{Hyperparameter Sensitivity}
In this section we discuss the impact of the hyperparameters $\epsilon$ and $c$ on FedRight.

When generating adversarial examples, the hyperparameter $\epsilon$ is used to limit the extracted intrinsic features of the model. We explore the effect of hyperparameters on property rights protection by setting $\epsilon$=0.01, $\epsilon$=0.08, and $\epsilon$=0.8.
The resutls are shown in Table.~\ref{tab:epsilonsensitivity}.
We can see the impact from $\epsilon$ is limited on FedRight performance.

\begin{table}[h]
\centering
\caption{Hyperparameter sensitivity analysis of $\epsilon$=0.01, $\epsilon$=0.08, and $\epsilon$=0.80.}\label{tab:epsilonsensitivity}
\resizebox{0.7\linewidth}{!}{ \tiny
\begin{tabular}{cl|ccc}
\hline \hline
\textbf{}    & \multicolumn{1}{c|}{\textbf{}} & \textbf{~~LeNet-5~~} & \textbf{~~VGG-11~~} & \textbf{~~ResNet-18~~} \\ \hline
\textbf{}    & \textbf{$\epsilon$=0.01}                & 100              & 100             & 100                \\
\textbf{DMC} & \textbf{$\epsilon$=0.08}                 & 100              & 100             & 100                \\
\textbf{}    & \textbf{$\epsilon$=0.80}                   & 100              & 100             & 100                \\ \hline \hline
\end{tabular}}
\end{table}

Regarding the sensitivity of selecting the specific number of key sample classes $c$, 
We test under three models, LeNet-5, VGG-11 and ResNet-18, setting $c$ = 5, 7, and 10,
the elevation rate of DMC$_{\rm Att}$ obtained in the face of PruningAtt compared to the average accuracy of random guesses (100/c).
As shown in Table~\ref{c2}, 
at PruningAtt is 0\%, it can be found that the number of classes does not seriously affect the validity of the IP validation. However, the elevation rate of DMC$_{\rm Att}$ can be effectively improved by increasing $c$ compared with the average accuracy of random guesses.

\begin{table}[h]
\centering
\caption{Hyperparameter sensitivity analysis of $c$=5, 7 and 10.}\label{c2}
\resizebox{0.8\linewidth}{!}{ 
\begin{tabular}{cl|ccccccccc}
\hline \hline
\multicolumn{2}{c|}{\multirow{3}{*}{\textbf{PruningAtt}}} & \multicolumn{3}{c}{\textbf{}}               & \multicolumn{3}{c}{\textbf{Elevation rate(\%)}} & \multicolumn{3}{c}{\textbf{}}               \\
\multicolumn{2}{c|}{}                                        & \multicolumn{3}{c}{\textbf{LeNet-5}}        & \multicolumn{3}{c}{\textbf{VGG-11}}         & \multicolumn{3}{c}{\textbf{ResNet-18}}      \\
\multicolumn{2}{c|}{}                                        & \textbf{C=5} & \textbf{C=7} & \textbf{C=10} & \textbf{C=5} & \textbf{C=7} & \textbf{C=10} & \textbf{C=5} & \textbf{C=7} & \textbf{C=10} \\ \hline
\multicolumn{2}{c|}{\textbf{90\%}}                             & 0.40         & 0.77         & \textbf{0.79}          & 0.00         & \textbf{0.02}         & 0.00          & -0.05        & 0.02         & \textbf{0.47}          \\
\multicolumn{2}{c|}{\textbf{60\%}}                             & 0.80         & 0.86         & \textbf{0.90}          & 0.80         & 0.86         & \textbf{0.90}          & 0.74         & 0.76         & \textbf{0.84}          \\
\multicolumn{2}{c|}{\textbf{30\%}}                             & 0.80         & 0.86         & \textbf{0.90}          & 0.80         & 0.86         & \textbf{0.90}          & 0.80         & 0.86         & \textbf{0.90}          \\
\multicolumn{2}{c|}{\textbf{0\%}}                              & 0.80         & 0.86         & \textbf{0.90}          & 0.80         & 0.86         & \textbf{0.90}          & 0.80         & 0.86         & \textbf{0.90}          \\ \hline \hline
\end{tabular}}
\end{table}

\begin{center}
\fcolorbox{black}{white!20}{\parbox{0.97\linewidth}
{
\emph{\textbf{Answer to RQ5:}}
The effectiveness of FedRight is not affected by the number of clients, data distribution and hyperparameter values. 
}
}
\end{center}

\section{Discussion}

With the widespread use of federated learning, the federated models are becoming more and more commercially valuable, and how to protect the property rights of federated models and prevent illegal profits after copying and stealing has practical significance. Our proposed FedRight framework can be effectively applied to several real-world scenarios, such as news recommendation systems, which have high business value and people benefit from personalized push messages from the recommendation system to meet their needs. In the FedRight framework, each user is a participant and the central server is the initiator of the IP declaration, which dynamically trains the detector according to the changes of the recommender system to achieve property rights protection.

\section{Conclusion\label{Conclusion}}
This paper presents FedRight, the FL copyright protection technique based on model fingerprints.
FedRight uses the features of the model to generate the model fingerprints with adaptive enhancement.
Using the distribution features of the model fingerprint input by the model, a detector is trained to claim the ownership of a suspect model.
FedRight does not require any training data, which is especially designed for FL scenario.
Moreover, FedRight only requires the input and output of the suspect model to verify the copyright, which fits the practical situation where model owner does not have access to the deployed stolen model. 
We extensively evaluate FedRight on nine models on three benchmark datasets to demostrate that it achieves remarkable results in terms of fidelity, effectiveness, and robustness.
To the best of our knowledge, we are the first to apply model fingerprints techniques to FL.
In future work, we plan to adapt FedRight to other high-value models such as speech recognition, and extend to other forms of deep learning architectures, such as recurrent neural networks.

It is interesting and challenging to verify the attribution of the model by comparing the features of the suspicious model and the original model, and it has important implications for future research, which we will investigate in our future work. 
Meanwhile, more comprehensively consider the case where the client is the initiator of property protection and the server has the possibility of being maliciously controlled.

\section*{Acknowledgements}
This research is supported by 
the National Natural Science Foundation of China (No. 62072406), 
Zhejiang Provincial Natural Science Foundation (No.LDQ23F020001),		
Chinese National Key Laboratory of Science and Technology on Information System Security (No. 61421110502), and 
National Key R\&D Projects of China (No. 2018AAA0100801).


% \bibliographystyle{model2-names}
% \bibliography{mybibfile.bib}

\begin{thebibliography}{10}
\providecommand{\url}[1]{#1}
\csname url@samestyle\endcsname
\providecommand{\newblock}{\relax}
\providecommand{\bibinfo}[2]{#2}
\providecommand{\BIBentrySTDinterwordspacing}{\spaceskip=0pt\relax}
\providecommand{\BIBentryALTinterwordstretchfactor}{4}
\providecommand{\BIBentryALTinterwordspacing}{\spaceskip=\fontdimen2\font plus
\BIBentryALTinterwordstretchfactor\fontdimen3\font minus
  \fontdimen4\font\relax}
\providecommand{\BIBforeignlanguage}[2]{{%
\expandafter\ifx\csname l@#1\endcsname\relax
\typeout{** WARNING: IEEEtran.bst: No hyphenation pattern has been}%
\typeout{** loaded for the language `#1'. Using the pattern for}%
\typeout{** the default language instead.}%
\else
\language=\csname l@#1\endcsname
\fi
#2}}
\providecommand{\BIBdecl}{\relax}
\BIBdecl

\bibitem{mcmahan2017communication}
\BIBentryALTinterwordspacing
B.~McMahan, E.~Moore, D.~Ramage, S.~Hampson, and B.~A. y~Arcas,
  ``Communication-efficient learning of deep networks from decentralized
  data,'' in \emph{Proceedings of the 20th International Conference on
  Artificial Intelligence and Statistics, {AISTATS} 2017, 20-22 April 2017,
  Fort Lauderdale, FL, {USA}}, ser. Proceedings of Machine Learning Research,
  vol.~54.\hskip 1em plus 0.5em minus 0.4em\relax {PMLR}, 2017, pp. 1273--1282.
  [Online]. Available: \url{http://proceedings.mlr.press/v54/mcmahan17a.html}
\BIBentrySTDinterwordspacing

\bibitem{mcmahan2016federated}
\BIBentryALTinterwordspacing
H.~B. McMahan, E.~Moore, D.~Ramage, and B.~A. y~Arcas, ``Federated learning of
  deep networks using model averaging,'' \emph{CoRR}, vol. abs/1602.05629,
  2016. [Online]. Available: \url{http://arxiv.org/abs/1602.05629}
\BIBentrySTDinterwordspacing

\bibitem{yang2019federated}
\BIBentryALTinterwordspacing
Q.~Yang, Y.~Liu, T.~Chen, and Y.~Tong, ``Federated machine learning: Concept
  and applications,'' \emph{{ACM} Trans. Intell. Syst. Technol.}, vol.~10,
  no.~2, pp. 12:1--12:19, 2019. [Online]. Available:
  \url{https://doi.org/10.1145/3298981}
\BIBentrySTDinterwordspacing

\bibitem{Justicia21}
\BIBentryALTinterwordspacing
A.~Blanco{-}Justicia, J.~Domingo{-}Ferrer, S.~Mart{\'{\i}}nez,
  D.~S{\'{a}}nchez, A.~Flanagan, and K.~E. Tan, ``Achieving security and
  privacy in federated learning systems: Survey, research challenges and future
  directions,'' \emph{Eng. Appl. Artif. Intell.}, vol. 106, p. 104468, 2021.
  [Online]. Available: \url{https://doi.org/10.1016/j.engappai.2021.104468}
\BIBentrySTDinterwordspacing

\bibitem{journals/isci/JiangXZ21}
\BIBentryALTinterwordspacing
C.~Jiang, C.~Xu, and Y.~Zhang, ``{PFLM:} privacy-preserving federated learning
  with membership proof,'' \emph{Inf. Sci.}, vol. 576, pp. 288--311, 2021.
  [Online]. Available: \url{https://doi.org/10.1016/j.ins.2021.05.077}
\BIBentrySTDinterwordspacing

\bibitem{WangZLZL21}
\BIBentryALTinterwordspacing
F.~Wang, H.~Zhu, R.~Lu, Y.~Zheng, and H.~Li, ``A privacy-preserving and
  non-interactive federated learning scheme for regression training with
  gradient descent,'' \emph{Inf. Sci.}, vol. 552, pp. 183--200, 2021. [Online].
  Available: \url{https://doi.org/10.1016/j.ins.2020.12.007}
\BIBentrySTDinterwordspacing

\bibitem{fd}
\BIBentryALTinterwordspacing
C.~Zhang, Y.~Xie, H.~Bai, B.~Yu, W.~Li, and Y.~Gao, ``A survey on federated
  learning,'' \emph{Knowl. Based Syst.}, vol. 216, p. 106775, 2021. [Online].
  Available: \url{https://doi.org/10.1016/j.knosys.2021.106775}
\BIBentrySTDinterwordspacing

\bibitem{conf/icdm/Shingi20}
\BIBentryALTinterwordspacing
G.~Shingi, ``A federated learning based approach for loan defaults
  prediction,'' in \emph{20th International Conference on Data Mining
  Workshops, {ICDM} Workshops 2020, Sorrento, Italy, November 17-20, 2020},
  G.~D. Fatta, V.~S. Sheng, A.~Cuzzocrea, C.~Zaniolo, and X.~Wu, Eds.\hskip 1em
  plus 0.5em minus 0.4em\relax {IEEE}, 2020, pp. 362--368. [Online]. Available:
  \url{https://doi.org/10.1109/ICDMW51313.2020.00057}
\BIBentrySTDinterwordspacing

\bibitem{KuoP22}
\BIBentryALTinterwordspacing
T.~Kuo and A.~Pham, ``Detecting model misconducts in decentralized healthcare
  federated learning,'' \emph{Int. J. Medical Informatics}, vol. 158, no.
  February, p. 104658, 2022. [Online]. Available:
  \url{https://doi.org/10.1016/j.ijmedinf.2021.104658}
\BIBentrySTDinterwordspacing

\bibitem{rs}
\BIBentryALTinterwordspacing
O.~A. Wahab, G.~Rjoub, J.~Bentahar, and R.~Cohen, ``Federated against the cold:
  {A} trust-based federated learning approach to counter the cold start problem
  in recommendation systems,'' \emph{Inf. Sci.}, vol. 601, pp. 189--206, 2022.
  [Online]. Available: \url{https://doi.org/10.1016/j.ins.2022.04.027}
\BIBentrySTDinterwordspacing

\bibitem{UchidaNSS17}
\BIBentryALTinterwordspacing
Y.~Uchida, Y.~Nagai, S.~Sakazawa, and S.~Satoh, ``Embedding watermarks into
  deep neural networks,'' in \emph{Proceedings of the 2017 {ACM} on
  International Conference on Multimedia Retrieval, {ICMR} 2017, Bucharest,
  Romania, June 6-9, 2017}.\hskip 1em plus 0.5em minus 0.4em\relax {ACM}, 2017,
  pp. 269--277. [Online]. Available:
  \url{https://doi.org/10.1145/3078971.3078974}
\BIBentrySTDinterwordspacing

\bibitem{Vybornova21}
\BIBentryALTinterwordspacing
Y.~D. Vybornova, ``Method for copyright protection of deep neural networks
  using digital watermarking,'' in \emph{Fourteenth International Conference on
  Machine Vision, {ICMV} 2021, Rome, Italy, November 8-12, 2021}, ser. {SPIE}
  Proceedings, vol. 12084.\hskip 1em plus 0.5em minus 0.4em\relax {SPIE}, 2021,
  p. 1208412. [Online]. Available: \url{https://doi.org/10.1117/12.2623444}
\BIBentrySTDinterwordspacing

\bibitem{LiZZDZX20}
\BIBentryALTinterwordspacing
M.~Li, Q.~Zhong, L.~Y. Zhang, Y.~Du, J.~Zhang, and Y.~Xiang, ``Protecting the
  intellectual property of deep neural networks with watermarking: The
  frequency domain approach,'' in \emph{19th {IEEE} International Conference on
  Trust, Security and Privacy in Computing and Communications, TrustCom 2020,
  Guangzhou, China, December 29, 2020 - January 1, 2021}.\hskip 1em plus 0.5em
  minus 0.4em\relax {IEEE}, 2020, pp. 402--409. [Online]. Available:
  \url{https://doi.org/10.1109/TrustCom50675.2020.00062}
\BIBentrySTDinterwordspacing

\bibitem{GuoP18}
\BIBentryALTinterwordspacing
J.~Guo and M.~Potkonjak, ``Watermarking deep neural networks for embedded
  systems,'' in \emph{Proceedings of the International Conference on
  Computer-Aided Design, {ICCAD} 2018, San Diego, CA, USA, November 05-08,
  2018}.\hskip 1em plus 0.5em minus 0.4em\relax {ACM}, 2018, p. 133. [Online].
  Available: \url{https://doi.org/10.1145/3240765.3240862}
\BIBentrySTDinterwordspacing

\bibitem{LiHZG19}
\BIBentryALTinterwordspacing
Z.~Li, C.~Hu, Y.~Zhang, and S.~Guo, ``How to prove your model belongs to you: a
  blind-watermark based framework to protect intellectual property of {DNN},''
  in \emph{Proceedings of the 35th Annual Computer Security Applications
  Conference, {ACSAC} 2019, San Juan, PR, USA, December 09-13, 2019}.\hskip 1em
  plus 0.5em minus 0.4em\relax {ACM}, 2019, pp. 126--137. [Online]. Available:
  \url{https://doi.org/10.1145/3359789.3359801}
\BIBentrySTDinterwordspacing

\bibitem{ZhaoHLMCH20}
\BIBentryALTinterwordspacing
J.~Zhao, Q.~Hu, G.~Liu, X.~Ma, F.~Chen, and M.~M. Hassan, ``{AFA:} adversarial
  fingerprinting authentication for deep neural networks,'' \emph{Comput.
  Commun.}, vol. 150, pp. 488--497, 2020. [Online]. Available:
  \url{https://doi.org/10.1016/j.comcom.2019.12.016}
\BIBentrySTDinterwordspacing

\bibitem{LukasZK21}
\BIBentryALTinterwordspacing
N.~Lukas, Y.~Zhang, and F.~Kerschbaum, ``Deep neural network fingerprinting by
  conferrable adversarial examples,'' in \emph{9th International Conference on
  Learning Representations, {ICLR} 2021, Virtual Event, Austria, May 3-7,
  2021}.\hskip 1em plus 0.5em minus 0.4em\relax OpenReview.net, 2021. [Online].
  Available: \url{https://openreview.net/forum?id=VqzVhqxkjH1}
\BIBentrySTDinterwordspacing

\bibitem{CaoJG21}
\BIBentryALTinterwordspacing
X.~Cao, J.~Jia, and N.~Z. Gong, ``Ipguard: Protecting intellectual property of
  deep neural networks via fingerprinting the classification boundary,'' in
  \emph{{ASIA} {CCS} '21: {ACM} Asia Conference on Computer and Communications
  Security, Virtual Event, Hong Kong, June 7-11, 2021}.\hskip 1em plus 0.5em
  minus 0.4em\relax {ACM}, 2021, pp. 14--25. [Online]. Available:
  \url{https://doi.org/10.1145/3433210.3437526}
\BIBentrySTDinterwordspacing

\bibitem{xue2020intellectual}
M.~Xue, Y.~Zhang, J.~Wang, and W.~Liu, ``Intellectual property protection for
  deep learning models: Taxonomy, methods, attacks, and evaluations,'' 2020.

\bibitem{Boenisch21}
\BIBentryALTinterwordspacing
F.~Boenisch, ``A systematic review on model watermarking for neural networks,''
  \emph{Frontiers Big Data}, vol.~4, p. 729663, 2021. [Online]. Available:
  \url{https://doi.org/10.3389/fdata.2021.729663}
\BIBentrySTDinterwordspacing

\bibitem{waffle}
\BIBentryALTinterwordspacing
B.~G.~A. Tekgul, Y.~Xia, S.~Marchal, and N.~Asokan, ``{WAFFLE:} watermarking in
  federated learning,'' in \emph{40th International Symposium on Reliable
  Distributed Systems, {SRDS} 2021, Chicago, IL, USA, September 20-23,
  2021}.\hskip 1em plus 0.5em minus 0.4em\relax {IEEE}, 2021, pp. 310--320.
  [Online]. Available: \url{https://doi.org/10.1109/SRDS53918.2021.00038}
\BIBentrySTDinterwordspacing

\bibitem{ms}
\BIBentryALTinterwordspacing
F.~Li, S.~Wang, and A.~W. Liew, ``Watermarking protocol for deep neural network
  ownership regulation in federated learning,'' in \emph{{IEEE} International
  Conference on Multimedia and Expo Workshops, {ICME} Workshops 2022, Taipei,
  Taiwan, July 18-22, 2022}.\hskip 1em plus 0.5em minus 0.4em\relax {IEEE},
  2022, pp. 1--4. [Online]. Available:
  \url{https://doi.org/10.1109/ICMEW56448.2022.9859395}
\BIBentrySTDinterwordspacing

\bibitem{fedipr}
\BIBentryALTinterwordspacing
L.~Fan, B.~Li, H.~Gu, J.~Li, and Q.~Yang, ``Fedipr: Ownership verification for
  federated deep neural network models,'' \emph{CoRR}, vol. abs/2109.13236,
  2021. [Online]. Available: \url{https://arxiv.org/abs/2109.13236}
\BIBentrySTDinterwordspacing

\bibitem{LuoLWX18}
\BIBentryALTinterwordspacing
B.~Luo, Y.~Liu, L.~Wei, and Q.~Xu, ``Towards imperceptible and robust
  adversarial example attacks against neural networks,'' in \emph{Proceedings
  of the Thirty-Second {AAAI} Conference on Artificial Intelligence, (AAAI-18),
  the 30th innovative Applications of Artificial Intelligence (IAAI-18), and
  the 8th {AAAI} Symposium on Educational Advances in Artificial Intelligence
  (EAAI-18), New Orleans, Louisiana, USA, February 2-7, 2018}.\hskip 1em plus
  0.5em minus 0.4em\relax {AAAI} Press, 2018, pp. 1652--1659. [Online].
  Available:
  \url{https://www.aaai.org/ocs/index.php/AAAI/AAAI18/paper/view/16217}
\BIBentrySTDinterwordspacing

\bibitem{HuangPGDA17}
\BIBentryALTinterwordspacing
S.~H. Huang, N.~Papernot, I.~J. Goodfellow, Y.~Duan, and P.~Abbeel,
  ``Adversarial attacks on neural network policies,'' in \emph{5th
  International Conference on Learning Representations, {ICLR} 2017, Toulon,
  France, April 24-26, 2017, Workshop Track Proceedings}.\hskip 1em plus 0.5em
  minus 0.4em\relax OpenReview.net, 2017. [Online]. Available:
  \url{https://openreview.net/forum?id=ryvlRyBKl}
\BIBentrySTDinterwordspacing

\bibitem{ZhengZGLP20}
\BIBentryALTinterwordspacing
H.~Zheng, Z.~Zhang, J.~Gu, H.~Lee, and A.~Prakash, ``Efficient adversarial
  training with transferable adversarial examples,'' in \emph{2020 {IEEE/CVF}
  Conference on Computer Vision and Pattern Recognition, {CVPR} 2020, Seattle,
  WA, USA, June 13-19, 2020}.\hskip 1em plus 0.5em minus 0.4em\relax Computer
  Vision Foundation / {IEEE}, 2020, pp. 1178--1187. [Online]. Available:
  \url{https://openaccess.thecvf.com/content\_CVPR\_2020/html/Zheng\_Efficient\_Adversarial\_Training\_With\_Transferable\_Adversarial\_Examples\_CVPR\_2020\_paper.html}
\BIBentrySTDinterwordspacing

\bibitem{WeiLLL018}
\BIBentryALTinterwordspacing
L.~Wei, B.~Luo, Y.~Li, Y.~Liu, and Q.~Xu, ``I know what you see: Power
  side-channel attack on convolutional neural network accelerators,'' in
  \emph{Proceedings of the 34th Annual Computer Security Applications
  Conference, {ACSAC} 2018, San Juan, PR, USA, December 03-07, 2018}.\hskip 1em
  plus 0.5em minus 0.4em\relax {ACM}, 2018, pp. 393--406. [Online]. Available:
  \url{https://doi.org/10.1145/3274694.3274696}
\BIBentrySTDinterwordspacing

\bibitem{JiaCCP21}
\BIBentryALTinterwordspacing
H.~Jia, C.~A. Choquette{-}Choo, V.~Chandrasekaran, and N.~Papernot, ``Entangled
  watermarks as a defense against model extraction,'' in \emph{30th {USENIX}
  Security Symposium, {USENIX} Security 2021, August 11-13, 2021}.\hskip 1em
  plus 0.5em minus 0.4em\relax {USENIX} Association, 2021, pp. 1937--1954.
  [Online]. Available:
  \url{https://www.usenix.org/conference/usenixsecurity21/presentation/jia}
\BIBentrySTDinterwordspacing

\bibitem{HitajHM19}
\BIBentryALTinterwordspacing
D.~Hitaj, B.~Hitaj, and L.~V. Mancini, ``Evasion attacks against watermarking
  techniques found in mlaas systems,'' in \emph{6th International Conference on
  Software Defined Systems, {SDS} 2019, Rome, Italy, June 10-13, 2019}.\hskip
  1em plus 0.5em minus 0.4em\relax {IEEE}, 2019, pp. 55--63. [Online].
  Available: \url{https://doi.org/10.1109/SDS.2019.8768572}
\BIBentrySTDinterwordspacing

\bibitem{removal}
\BIBentryALTinterwordspacing
M.~Shafieinejad, J.~Wang, N.~Lukas, and F.~Kerschbaum, ``On the robustness of
  the backdoor-based watermarking in deep neural networks,'' \emph{CoRR}, vol.
  abs/1906.07745, 2019. [Online]. Available:
  \url{http://arxiv.org/abs/1906.07745}
\BIBentrySTDinterwordspacing

\bibitem{RouhaniCK19}
\BIBentryALTinterwordspacing
B.~D. Rouhani, H.~Chen, and F.~Koushanfar, ``Deepsigns: An end-to-end
  watermarking framework for ownership protection of deep neural networks,'' in
  \emph{Proceedings of the Twenty-Fourth International Conference on
  Architectural Support for Programming Languages and Operating Systems,
  {ASPLOS} 2019, Providence, RI, USA, April 13-17, 2019}.\hskip 1em plus 0.5em
  minus 0.4em\relax {ACM}, 2019, pp. 485--497. [Online]. Available:
  \url{https://doi.org/10.1145/3297858.3304051}
\BIBentrySTDinterwordspacing

\bibitem{NambaS19}
\BIBentryALTinterwordspacing
R.~Namba and J.~Sakuma, ``Robust watermarking of neural network with
  exponential weighting,'' in \emph{Proceedings of the 2019 {ACM} Asia
  Conference on Computer and Communications Security, AsiaCCS 2019, Auckland,
  New Zealand, July 09-12, 2019}.\hskip 1em plus 0.5em minus 0.4em\relax {ACM},
  2019, pp. 228--240. [Online]. Available:
  \url{https://doi.org/10.1145/3321705.3329808}
\BIBentrySTDinterwordspacing

\bibitem{fgsm}
\BIBentryALTinterwordspacing
I.~J. Goodfellow, J.~Shlens, and C.~Szegedy, ``Explaining and harnessing
  adversarial examples,'' in \emph{3rd International Conference on Learning
  Representations, {ICLR} 2015, San Diego, CA, USA, May 7-9, 2015, Conference
  Track Proceedings}, 2015. [Online]. Available:
  \url{http://arxiv.org/abs/1412.6572}
\BIBentrySTDinterwordspacing

\bibitem{cw}
\BIBentryALTinterwordspacing
N.~Carlini and D.~A. Wagner, ``Towards evaluating the robustness of neural
  networks,'' \emph{CoRR}, vol. abs/1608.04644, 2016. [Online]. Available:
  \url{http://arxiv.org/abs/1608.04644}
\BIBentrySTDinterwordspacing

\bibitem{pgd}
\BIBentryALTinterwordspacing
A.~Madry, A.~Makelov, L.~Schmidt, D.~Tsipras, and A.~Vladu, ``Towards deep
  learning models resistant to adversarial attacks,'' in \emph{6th
  International Conference on Learning Representations, {ICLR} 2018, Vancouver,
  BC, Canada, April 30 - May 3, 2018, Conference Track Proceedings}.\hskip 1em
  plus 0.5em minus 0.4em\relax OpenReview.net, 2018. [Online]. Available:
  \url{https://openreview.net/forum?id=rJzIBfZAb}
\BIBentrySTDinterwordspacing

\bibitem{pytorch}
\BIBentryALTinterwordspacing
A.~Sankaran, N.~A. Alashti, C.~Psarras, and P.~Bientinesi, ``Benchmarking the
  linear algebra awareness of tensorflow and pytorch,'' \emph{CoRR}, vol.
  abs/2202.09888, 2022. [Online]. Available:
  \url{https://arxiv.org/abs/2202.09888}
\BIBentrySTDinterwordspacing

\bibitem{mnist}
\BIBentryALTinterwordspacing
Y.~LeCun, L.~Bottou, Y.~Bengio, and P.~Haffner, ``Gradient-based learning
  applied to document recognition,'' \emph{Proc. {IEEE}}, vol.~86, no.~11, pp.
  2278--2324, 1998. [Online]. Available: \url{https://doi.org/10.1109/5.726791}
\BIBentrySTDinterwordspacing

\bibitem{cifar10}
\BIBentryALTinterwordspacing
M.~Ayi and M.~El{-}Sharkawy, ``Rmnv2: Reduced mobilenet {V2} for {CIFAR10},''
  in \emph{10th Annual Computing and Communication Workshop and Conference,
  {CCWC} 2020, Las Vegas, NV, USA, January 6-8, 2020}.\hskip 1em plus 0.5em
  minus 0.4em\relax {IEEE}, 2020, pp. 287--292. [Online]. Available:
  \url{https://doi.org/10.1109/CCWC47524.2020.9031131}
\BIBentrySTDinterwordspacing

\bibitem{LeNet5}
\BIBentryALTinterwordspacing
A.~El{-}Sawy, H.~M. El{-}Bakry, and M.~Loey, ``{CNN} for handwritten arabic
  digits recognition based on lenet-5,'' in \emph{Proceedings of the
  International Conference on Advanced Intelligent Systems and Informatics,
  {AISI} 2016, Cairo, Egypt, October 24-26, 2016}, ser. Advances in Intelligent
  Systems and Computing, vol. 533, 2016, pp. 566--575. [Online]. Available:
  \url{https://doi.org/10.1007/978-3-319-48308-5\_54}
\BIBentrySTDinterwordspacing

\bibitem{vgg16}
\BIBentryALTinterwordspacing
H.~Chen, ``Reliable and efficient distributed machine learning,'' Ph.D.
  dissertation, Royal Institute of Technology, Stockholm, Sweden, 2022.
  [Online]. Available:
  \url{https://nbn-resolving.org/urn:nbn:se:kth:diva-310374}
\BIBentrySTDinterwordspacing

\bibitem{resnet18}
\BIBentryALTinterwordspacing
K.~He, X.~Zhang, S.~Ren, and J.~Sun, ``Deep residual learning for image
  recognition,'' in \emph{2016 {IEEE} Conference on Computer Vision and Pattern
  Recognition, {CVPR} 2016, Las Vegas, NV, USA, June 27-30, 2016}.\hskip 1em
  plus 0.5em minus 0.4em\relax {IEEE} Computer Society, 2016, pp. 770--778.
  [Online]. Available: \url{https://doi.org/10.1109/CVPR.2016.90}
\BIBentrySTDinterwordspacing

\bibitem{hfl}
\BIBentryALTinterwordspacing
W.~Wu, ``Towards efficient horizontal federated learning,'' Ph.D. dissertation,
  University of Warwick, Coventry, {UK}, 2021. [Online]. Available:
  \url{https://ethos.bl.uk/OrderDetails.do?uin=uk.bl.ethos.856355}
\BIBentrySTDinterwordspacing

\bibitem{rm1}
\BIBentryALTinterwordspacing
M.~Shafieinejad, N.~Lukas, J.~Wang, X.~Li, and F.~Kerschbaum, ``On the
  robustness of backdoor-based watermarking in deep neural networks,'' in
  \emph{IH{\&}MMSec '21: {ACM} Workshop on Information Hiding and Multimedia
  Security, Virtual Event, Belgium, June, 22-25, 2021}.\hskip 1em plus 0.5em
  minus 0.4em\relax {ACM}, 2021, pp. 177--188. [Online]. Available:
  \url{https://doi.org/10.1145/3437880.3460401}
\BIBentrySTDinterwordspacing

\bibitem{rm2}
\BIBentryALTinterwordspacing
S.~Sun, H.~Wang, M.~Xue, Y.~Zhang, J.~Wang, and W.~Liu, ``Detect and remove
  watermark in deep neural networks via generative adversarial networks,'' in
  \emph{Information Security - 24th International Conference, {ISC} 2021,
  Virtual Event, November 10-12, 2021, Proceedings}, ser. Lecture Notes in
  Computer Science, vol. 13118.\hskip 1em plus 0.5em minus 0.4em\relax
  Springer, 2021, pp. 341--357. [Online]. Available:
  \url{https://doi.org/10.1007/978-3-030-91356-4\_18}
\BIBentrySTDinterwordspacing

\bibitem{rm3}
\BIBentryALTinterwordspacing
N.~Chattopadhyay, C.~S.~Y. Viroy, and A.~Chattopadhyay, ``Re-markable: Stealing
  watermarked neural networks through synthesis,'' in \emph{Security, Privacy,
  and Applied Cryptography Engineering - 10th International Conference, {SPACE}
  2020, Kolkata, India, December 17-21, 2020, Proceedings}, ser. Lecture Notes
  in Computer Science, vol. 12586.\hskip 1em plus 0.5em minus 0.4em\relax
  Springer, 2020, pp. 46--65. [Online]. Available:
  \url{https://doi.org/10.1007/978-3-030-66626-2\_3}
\BIBentrySTDinterwordspacing

\bibitem{cifar100}
\BIBentryALTinterwordspacing
S.~Singla, S.~Singla, and S.~Feizi, ``Improved deterministic l2 robustness on
  {CIFAR-10} and {CIFAR-100},'' in \emph{The Tenth International Conference on
  Learning Representations, {ICLR} 2022, Virtual Event, April 25-29,
  2022}.\hskip 1em plus 0.5em minus 0.4em\relax OpenReview.net, 2022. [Online].
  Available: \url{https://openreview.net/forum?id=tD7eCtaSkR}
\BIBentrySTDinterwordspacing

\end{thebibliography}

\end{document}